\documentclass[letterpaper]{aa520}
\usepackage{natbib}
\bibpunct{(}{)}{;}{a}{}{,}
\usepackage{psfig}
\usepackage{minipage}
\usepackage{epsfig,afterpage} 
\usepackage{graphicx}
\usepackage{pifont}
\usepackage{amssymb}
\usepackage{longtable}
\usepackage{array}
\usepackage{amsmath}
\usepackage{rotating}
\usepackage{multirow}
\usepackage{lscape}

\begin{document}

\title {The Age--Activity--Rotation Relationship in Solar--Type Stars.
\thanks{Observations collected at the ESO VLT. Some data presented herein were 
obtained at the W. M. Keck Observatory, which is operated as a scientific 
partnership among the California Institute of Technology, the University 
of California, and the National Aeronautics and Space Administration. The 
Observatory was made possible by the generous financial support of the W. 
M. Keck Foundation. }}
  \subtitle{ }

\author{ G. Pace\inst{1,2} \and L. Pasquini\inst{1}}

\offprints{G. Pace, \email gpace@eso.org}

\institute{ European Southern Observatory, 
            Karl Schwarzschild Strasse 2, 85748, 
            Garching bei M\"{u}nchen, Germany \\ 
       \and
            Dipartimento di Astronomia, Universit\`a di Trieste, 
            Via G.B. Tiepolo 11, 34131 Trieste, Italy \\
          }
\date{}

\abstract{
Measured from high--resolution  spectra, we present Ca $\!{\rm II}$\  K line chromospheric fluxes in 35
G dwarf stars of 5 open clusters to determine the age--activity--rotation relationship
from the young Hyades and 
Praesepe  (0.6 Gyr) to the old M 67 ($\sim$ 4.5 Gyr) through the two 
intermediate age clusters IC 4651 and NGC 3680 ($\sim$ 1.7 Gyr).
The full amplitude of the activity index within a cluster is
slightly above 60 \%  for every cluster but one,  NGC 3680, in which
only two stars were observed. As a comparison, the same  Solar 
Ca $\!{\rm II}$\  index  varies by $\sim 40 \%$ during a solar cycle.
Four of our clusters (Hyades and Praesepe, IC 4651 and NGC 3680) are couple of 
twins as far as age is concerned; the Hyades  have the same chromospheric--activity 
level than Praesepe, at odds 
with early claims based on X--ray observations.
Both stars in NGC 3680 are indistinguishable,  as far as chromospheric activity is concerned, 
from those in the coeval IC 4651. This is a nice validation of the existence of an age--activity relationship.
On the other hand, the two intermediate age clusters have the same activity level of 
the much older M 67 and of the Sun. Our data therefore shows that a dramatic decrease in
chromospheric activity takes place among solar stars between the Hyades and the 
IC4651 age, in about  1 Gyr. Afterwards,  activity remains
virtually constant for more than 3 Gyr. 
We have also measured $v \sin i$ for all of our stars and the average rotational velocity shows the same trend 
as the chromospheric--activity index. 
We briefly investigate the impact of this result on the age determinations of  field G dwarfs in 
the solar neighborhood; the two main 
conclusions are that a consistent group of 'young' stars (i.e. as active as Hyades stars) is present,
and that it is virtually impossible to give accurate chromospheric ages for 
stars older than $\sim$ 2 Gyr. The observed abrupt decline in activity explains very well the Vaughan-Preston gap.
      \keywords{Stars: late--type -- Stars: activity -- Stars: chromosphere -- Stars: rotation}}

\titlerunning{The Age--Activity--Rotation Relationship.}

\maketitle

\section{Introduction}
Chromospheric activity is powered by rotation in the 
presence of a deep convective envelope. Because of rotation braking by a magnetised stellar wind,
rotational velocity decreases with age, as well as activity, unless the 
angular momentum is sustained by tidal interaction, as in the 
case of short--period binaries.
This simple picture is at the basis of the so called age--activity--rotation relationship
in solar--type stars and forms the paradigm of stellar activity in the 
last 30 years. The literature on the argument is enormous and the reader may refer to a number of works,
among which we mention the  Parker's review on the solar magnetic fields \citep{parker70}.

The present study aims at improving this picture by accurately ascertaining the form of the activity--age law 
on the basis of a new dataset. 
The decaying law of  chromospheric activity is
a precious tool in determining ages of field main sequence stars and 
chromospheric ages have been used to undertake 
detailed studies of the star formation rate 
and chemical enrichment of the Galaxy  \citep[]{rmsf00b,rmsf00a}. 

The flux emission in the core of  the K line of  Ca $\!{\rm II}$\ is the most effective tracer to probe the  
chromospheric activity in solar--type stars  \citep[see e.g.][]{bcmr74,la78}. 
Indeed, we intend to use in this study the equivalent width of the line emission core, corrected for the 
underlying stellar photospheric contribution and transformed into flux at the stellar surface ($F_K^{\rm \prime}$), 
as index of the level of chromospheric activity \citep{ppp88}.

Among similar indices the 
most used and widely known is the Mount Wilson $R_{HK}^{\rm \prime}$, which is the flux in the 
Mount Wilson photometer passband normalized to the 
stellar effective temperature \citep{nhbdv84}. The apex  indicates that the quantity is corrected for the 
the photospheric--flux contribution. This convention will be adopted throughout this paper.
While the Mount Wilson's index is a contrast index, our  $F_K^{\rm \prime}$ is an absolute flux.
If we indicate the chromospheric flux of the H and K line cores at the stellar 
surface (erg cm$^{-2 }$sec$^{-1}$) as $F_{HK}^{\rm \prime}$, and the bolometric one as ${F_{BOL}}$,
then $R_{HK}^{\rm \prime}$ is the ratio $\frac{F_{HK}^{\rm \prime}}{F_{BOL}}$.

While chromospheric activity has been widely studied in field stars 
\citep[see e.g.][]{hsdb96,w63,pbp90,luca92},
relatively little work has been done on clusters, which are the perfect 
calibrators of the age--activity--rotation relationship: they are in fact homogeneous samples of 
stars with well determined ages. 

For the reasons explained in \citet{vBm80}, old open clusters are rare and tend to be distant. 
This is why the original \citet{skumanich} 
law was based only on 1 point older than the Hyades, which was the Sun. 
Only with the advent of 4--m telescopes and  high--sensitivity instruments has it been possible to acquire 
Ca $\!{\rm II}$\  spectra for  subgiants in M 67 \citep{dwp99}, which is as distant as 850 pc \citep[see e.g.][]{cgbc96}. 
Nowadays 8--m class telescopes are sufficiently sensitive to allow the study of Ca $\!{\rm II}$\  emission even in the 
dwarfs of distant clusters.

\citet[]{bch87} made the most consistent attempt to cover the lack of
old--intermediate age clusters data.
They built a 2--\AA\  resolution system and used it to observe several intermediate and old clusters.
This very interesting attempt, on the other hand, needs to be looked with 
some caution, since, for instance, the typical G star emission core is much larger than 2 \AA. Therefore 
a  2--\AA\  resolution is sensitive to other factors, such as zero point corrections and 
photospheric profile in addition to activity \citep[see also][for a discussion on Barry et al.'s data]{sdj91}. 
A narrower, pure chromospheric--activity index based on high--resolution spectroscopy, such as 
the quoted  $F_K^{\rm \prime}$, is of course much more reliable.
This is the main ground of the present work.

\section{Observations and data sample}
\label{data}
Our study is based on spectra of 35 stars in 5 open clusters and on the solar data from \citet{wl81}.
White and Livingston made a very detailed study of the Ca $\!{\rm II}$\   solar K line covering the cycle 21 from 
the first minimum to the maximum. Because of differences in data analysis and reduction, their 1--\AA\ indices
have to be multiplied by 1.11 to be homogeneous with ours.

The Praesepe, NGC 3680, IC 4651 and M 67 observations were obtained with the 
UVES spectrograph at the VLT Kueyen telescope. The blue slit was set at 0.7 
arc seconds giving a resolving power of R $\sim$ 60000 \citep{ddkdk00}.
For the faintest stars, up to 3 consecutive 90--minutes exposures were co added, in order to reach 
enough signal at the bottom of the deep Ca $\!{\rm II}$\  K line which is typically less than 
10$\%$ of the continuum in non active stars. 
The signal to noise ratio per pixel at the bottom of the K line ranges from about 10 for the worst 
spectra in M 67 to about 30 for most of the spectra in Praesepe.

The red slit was set to 0.4 arc seconds, to obtain a resolution of R=100000

The data were reduced using the UVES pipeline \citep{bmbchkw00}, radial velocity corrected and finally co--added. 

The Hyades spectra were obtained within the program of search for extra--solar planets among the
solar--type stars of this cluster \citep{chp02,psch02} with HIRES at the Keck Observatory.
They have a resolution similar to the blue
UVES ones (R=60000), and a signal to noise ratio at the bottom of the K line ranging
from about 20 to about 30.  
The original Hyades' sample comprises a large range of spectral types, and was selected to 
include only stars with  $v \sin i$ lower than 15 km/sec. We have selected for the present study only the 
dwarfs with B$-$V colours similar to the solar one.
According to \citet{swgsw2000} (see the {\textit g} panel in their Figure 4), late--type stars which 
rotate faster than 15 km/sec are rare exceptions \citep[unless we are dealing with clusters much younger than the Hyades, see][]{bouvier}, 
therefore we do not expect to have a strong bias towards slow rotators.

All the Ca $\!{\rm II}$\  data have been normalized to the 3950.5 {\AA}  pseudocontinuum \citep{catalano78}.

The clusters span a wide range of ages, from the young Hyades and Praesepe (0.6 Gyr) to the old M 67 (4.8 Gyr).
Together with the Sun, they constitute three pairs of objects at three different ages:
Hyades and Praesepe, IC 4651 and NGC 3680 (1.6 and 1.8 Gyr), and M 67 and the Sun (4.8 and 4.48 Gyr).
This allows to have a first check on the consistency of the age--activity--rotation relationship: 
if each pair has the same chromospheric--activity level, we will have a strong confirmation  that 
chromospheric activity is an age driven parameter.

In addition, the comparison between Hyades and Praesepe is extremely interesting, after the 
finding by \citet{rs95} that the solar stars 
belonging to these clusters  have different X--ray (coronal) emission, with the Hyades showing substantially higher
X--ray luminosities than Praesepe \citep[see][and reference therein]{bsr97}.

The ages of M 67, NGC 3680 and IC 4651 are taken from \citet{cc94}.
Hyades and Praesepe, with some differences between various authors, are often thought to be coeval
with ages of about 0.6 Gyr \citep[see e.g.][]{perryman,mermilliod,allen}.
For the solar age we have used Guenther's work (1989), based on the age of the oldest meteorites.

All the stars observed with the VLT 
were chosen to be on the main sequence, high--probability members
of the clusters and not known to be binaries at the moment of the observation.
The choice was done using as references \citet{naa96} for NGC 3680, 
\citet{man02} for IC4651 and several sources for M 67, including \citet{lmmd92}
for the binary determination in this cluster. Our stars have colours comprised 
within 0.51 $<$(B$-$V)$_0 <$ 0.73, closely encompassing the solar colour which, with some 
variations between different authors, is evaluated   around B$-$V=0.63 \citep[see e.g.][]{sv94}.
We also note that the paucity of stars observed in NGC 3680 is imposed by the fact that,
as shown by \citet{naa97}, this cluster has very few single G stars members left,
most of its low mass members being dispersed in its lifetime. 
  
As far as Praesepe is concerned, out of the 7 stars observed we chose 3 stars  
detected in X--ray by \citet{rs95} and 4  undetected, in order to 
check whether the
chromospheric activity was different between the two groups, 
as expected if chromospheric and coronal activity were strongly 
correlated. 

All the observed stars are summarized in Table \ref{maintab}. Star names are taken from \citet{eggen69}
for NGC 3680, \citet{AMC} and \citet{eggen71} for IC 4651, \citet{sand} for M 67, \citet{kw} for Praesepe 
and \citet{vanbueren} for the Hyades stars.
The photometry is taken from  the reference papers for IC 4651, from \citet{naa97} for NGC 3680, 
from Hipparcos Catalogue \citep{hipparcos} for Hyades, from \citet{js91} for Praesepe, 
with the exception of KW 368 which is taken from \citet{jc83}, and from \citet{mmj93} for M 67. 
For Hyades  no correction at all for interstellar (IS) extinction is applied \citep{noext}, 
and NGC 3680 colours are 
given already dereddened by \citet{naa97}. B$-$V colours of Praespe, M 67 and IC 4651  had to be corrected for colour excess. 
We used E(B$-$V)=0.05 for M 67 \citep[also taken from][]{mmj93}, E(B$-$V)=0.086 for IC 4651 \citep{AMC1} and  E(B$-$V)=0.02 for Praesepe \citep{noext}. 
The uncertainties in  reddening are not at all critical for the analysis performed. 

The photometric data retrieved from the literature are shown
in the second and third columns of Table \ref{maintab}(V magnitude and B$-$V colour index, respectively).
The different chromospheric--activity indices are reported in Table \ref{maintab} 
from the fourth to the sixth column and explained in the caption.

The quantity shown in the seventh Column is the Full Width Half Maximum (FWHM) of the cross correlation profile of the 
spectra. The cross correlation profiles were obtained as described in \citet{mpd01}:
we used as template a box--shaped mask, suitable for solar--type stars, which was kindly provided by Claudio Melo. 
We processed the red part of the red arm UVES--spectra, all rebinned at the same step (0.0174087 \AA). The 1$\sigma$ 
uncertainties of the FWHMs are of about $1 \%$.
 
The rotational velocities determined by us are given in the last column .
They are obtained from the FWHMs via the calibrations of Section \ref{rot}.

It is worth noticing
that what really matters for the present purpose is the evolution of the angular momentum of the stars,
we are therefore mostly interested in the differential $v \sin i$ values.

\begin{table*}
\begin{center}
\begin{tabular}{c c c c c c c c}
\hline
\hline
  STAR         &  V  &(B$-$V)$_0$&1--\AA\ index &1--\AA\  index & $F_K^{\rm \prime}$ & FWHM   &$v\sin i$  \\
               &      &          &non ISM corr  &ISM corr       &ISM corr            &        &           \\
               &[mag] &[mag]     &   \AA        & \AA           & {\tiny 10$^6$  [erg$\cdot$cm$^{-2} \cdot $sec$^{-1}$]}%
                                                                                     &[pixels]&[km/sec]   \\
\hline
\hline
\multicolumn{8}{c}{PRAESEPE}\\
\hline
KW  49         &10.65 &0.59    & 0.219  & 0.219  &  2.10 & 14.95  &   8.57    \\
KW 100         &10.57 &0.51    & 0.238  & 0.238  &  3.03 & 20.27  &  13.06    \\
KW 326         &11.20 &0.56    & 0.220  & 0.220  &  2.35 & 11.54  &   5.15    \\
KW 368         &11.30 &0.71    & 0.258  & 0.258  &  1.59 & 10.91  &   4.39    \\
KW 392         &10.78 &0.51    & 0.231  & 0.231  &  2.92 & 12.09  &   5.77    \\
KW 418         &10.51 &0.53    & 0.219  & 0.219  &  2.57 & 13.12  &   6.84    \\
			                              
\hline			                              
\multicolumn{8}{c}{HYADES}\\                          
\hline			                              
van Bueren   1 & 7.38 & 0.567  & 0.165  &  0.165 & 1.55  & 12.46  &   5.5     \\   
van Bueren   2 & 7.78 & 0.617  & 0.208  &  0.208 & 1.78  & 12.65  &   5.5     \\   
van Bueren  10 & 7.85 & 0.593  & 0.266  &  0.266 & 2.66  & 13.04  &   6.2     \\   
van Bueren  15 & 8.05 & 0.657  & 0.293  &  0.293 & 2.34  & 12.60  &   5.4     \\  
van Bueren  17 & 8.45 & 0.693  & 0.257  &  0.257 & 1.71  & 12.10  &   4.5     \\   
van Bueren  18 & 8.05 & 0.640  & 0.279  &  0.279 & 2.36  & 12.46  &   5.4     \\   
van Bueren  31 & 7.46 & 0.560  & 0.246  &  0.246 & 2.70  & 16.39  &  10.0     \\
van Bueren  49 & 8.22 & 0.599  & 0.192  &  0.192 & 1.71  & 11.28  &   2.8     \\   
van Bueren  52 & 7.79 & 0.599  & 0.233  &  0.233 & 2.20  & 13.12  &   6.5     \\   
van Bueren  65 & 7.41 & 0.537  & 0.187  &  0.187 & 2.04  & 15.42  &   8.8     \\
van Bueren  66 & 7.49 & 0.560  & 0.238  &  0.238 & 2.59  & 14.88  &   8.6     \\
van Bueren  73 & 7.83 & 0.609  & 0.255  &  0.255 & 2.38  & 13.35  &   6.8     \\   
van Bueren  88 & 7.78 & 0.550  & 0.213  &  0.213 & 2.32  & 11.68  &   1.0     \\   
van Bueren  97 & 7.90 & 0.631  & 0.264  &  0.264 & 2.28  & 12.70  &   5.4     \\
van Bueren 118 & 7.72 & 0.578  & 0.146  &  0.146 & 1.25  & 12.55  &   5.3     \\   
\hline			         	  		
\multicolumn{8}{c}{IC 4651}\\   	  		
\hline			        	  		
  AMC 1109     &14.534&0.599   & 0.093  &  0.102 &  0.63 & 10.20  &    3.39   \\
  AMC 2207     &14.527&0.595   & 0.073  &  0.085 &  0.44 & 10.03  &    3.11   \\
  AMC 4220     &14.955&0.695   & 0.113  &  0.133 &  0.68 &  9.96  &    2.99   \\
  AMC 4226     &14.645&0.624   & 0.079  &  0.093 &  0.48 &  9.78  &    2.67   \\
  Eggen 45     &14.27 &0.53    & 0.068  &  0.074 &  0.37 & 10.26  &    3.48   \\
\hline			           	       		       		        
\multicolumn{8}{c}{NGC 3680}\\     	       		       		        
\hline			           	       		       		        
  Eggen 60     &14.290&0.641   & 0.090  &  0.095 &  0.47 &  9.88  & 2.85      \\
  Eggen 70     &14.589&0.651   & 0.075  &  0.084 & 0.34  & 10.49  & 3.82      \\
\hline			       	       	 	               
\multicolumn{8}{c}{M 67}\\      	       	 	               
\hline			       	       	 	               
Sanders  746   &14.380&0.659   & 0.091  &  0.108 &  0.56 &  9.45  & 1.96      \\
Sanders 1012   &14.516&0.687   & 0.116  &  0.133 &  0.71 &  9.66  & 2.43      \\
Sanders 1048   &14.411&0.635   & 0.073  &  0.085 &  0.37 &  9.62  & 2.35      \\
Sanders 1092   &13.308&0.581   & 0.074  &  0.088 &  0.50 & 10.99  & 4.49      \\
Sanders 1255   &14.486&0.616   & 0.086  &  0.105 &  0.63 &  9.80  & 2.71      \\
Sanders 1283   &14.115&0.590   & 0.067  &  0.082 &  0.41 & 10.25  & 3.46      \\
Sanders 1287   &14.030&0.558   & 0.079  &  0.090 &  0.56 & 10.39  & 3.67      \\
\hline
\end{tabular}
\end{center}
\caption{In this table all of the measurements regarding our sample stars are shown. 
The names in Column 1 are taken from \citet{eggen69} for NGC 3680, \citet{AMC} (AMC) and \citet{eggen71} 
for IC 4651, \citet{sand} for M 67, \citet{kw} for Praesepe and \citet{vanbueren}  
for the Hyades stars.
The photometry is taken from  the reference papers for IC 4651, from \citet{naa97} for NGC 3680, 
from Hipparcos Catalogue \citep{hipparcos} for Hyades, from \citet{js91} for Praesepe, 
with the exception of KW 368 which is taken from \citet{jc83}, and from \citet{mmj93} for M 67.
Regarding  chromospheric--activity indices measurements, there are the 1--\AA\ indices 
derived from spectra either non corrected (Columns 4) or corrected (Columns 5)
for the absorption of the IS medium. 
To the corrected index we have subtracted the photospheric contribution, which is
0.049 \AA, and transformed the result into flux. The final data are in Column 6.
$v \sin i$ measurements in Column 8 are taken from \citet{psc03} for Hyades, for all the other clusters
are  based on  the FWHMs in Column 7 via the calibrations in Section \ref{rot}.}
\label{maintab}
\end{table*}

\section{Analysis}

\subsection{Ca $\!{\rm II}$\ fluxes}
\label{fluxes}

The analysis is based on the measurement of flux in the emission core of the Ca $\!{\rm II}$\ K line. 
This line is the most sensitive indicator of chromospheric activity, and it has been widely studied 
in a number of different environments, including the Sun and solar stars \citep{la70,
lmrw79,ppp88,luca92,wl81}. 

In general, the most sensitive area is that comprised between the line K1 minima,  
but those are hardly seen in low activity stars or in intermediate S/N ratio 
spectra. 

A very useful index of the level of chromospheric activity is the area of the 1--\AA\  wide band
centered in the line core (hereafter referred to as 1--\AA\  index).
This 1--\AA\ band corresponds roughly to the separation between the K$_1$ minima for an active 
star (as those in Hyades or Praesepe), and it is about 30 $\%$ larger than the K$_1$ minima separation
for a star as active as the Sun \citep[]{wl81,luca92}.
The 1--\AA\ index was measured
in the  spectra either uncorrected or corrected for IS absorption.
The results are shown in the fourth and fifth column of  Table \ref{maintab} respectively. 
 In addition, we recall that the $F_K^{\rm \prime}$  is a measurement of a contrast 
between the chromospheric emission and the underlying photospheric flux. 
Several calibrations can be used to eliminate this offset and to quantify different 
bandwidths \citep{lmrw79}; this is widely discussed \citep[e.g. in][]{ppp88}, and 
will not be repeated here. 

In addition, although our clusters are among the closest and less reddened, the Ca $\!{\rm II}$\  lines could be 
contaminated by the IS Ca $\!{\rm II}$, a possibility that we have taken into account as 
explained later in this Section.

We summarize the different steps required to get the final fluxes from the observed spectra.

\begin{enumerate}
\item Spectra normalization and rectification
\item Subtraction of the IS line
\item Index measurement
\item Subtraction of the photospheric contribution
\item Calibration in Fluxes
\end{enumerate}

\begin{figure}
\resizebox{15cm}{8cm}{
\begin{minipage}{0.07\textwidth}
{\large $\frac{I}{I_{\lambda=3950.5}}$}
\end{minipage}
\begin{minipage}{0.9\textwidth}
\begin{tabular}{c}
{\includegraphics[height=16cm]{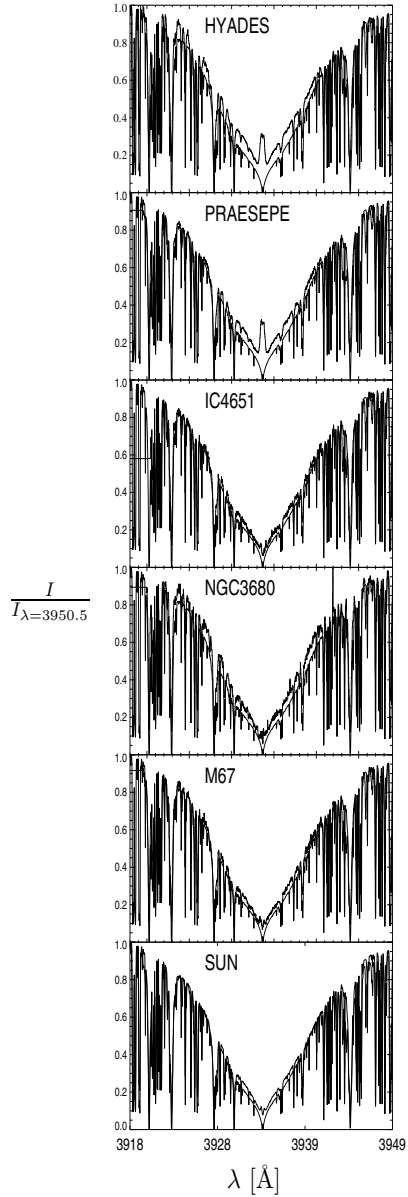}}\\
\\
{\large $\lambda$  [\AA]}\\
\end{tabular}
\end{minipage}
}
\caption{Average clusters Ca $\!{\rm II}$\   K line spectra. Overimposed is the synthetic photospheric--solar model. 
In the bottom panel we also plotted the real solar spectrum, also overimposed on its model.
The effect of the back heating of the chromosphere is evident also outside the core.}
\label{avspectra}
\end{figure}

\begin{figure*}
\resizebox{16cm}{8cm}{
\begin{minipage}{0.15\textwidth}
{\huge $\frac{I}{I_{\lambda=3950.5}}$}
\end{minipage}
\begin{minipage}{0.8\textwidth}
\begin{tabular}{c}
{\huge HYADES}\\
\\
{\includegraphics[height=16cm]{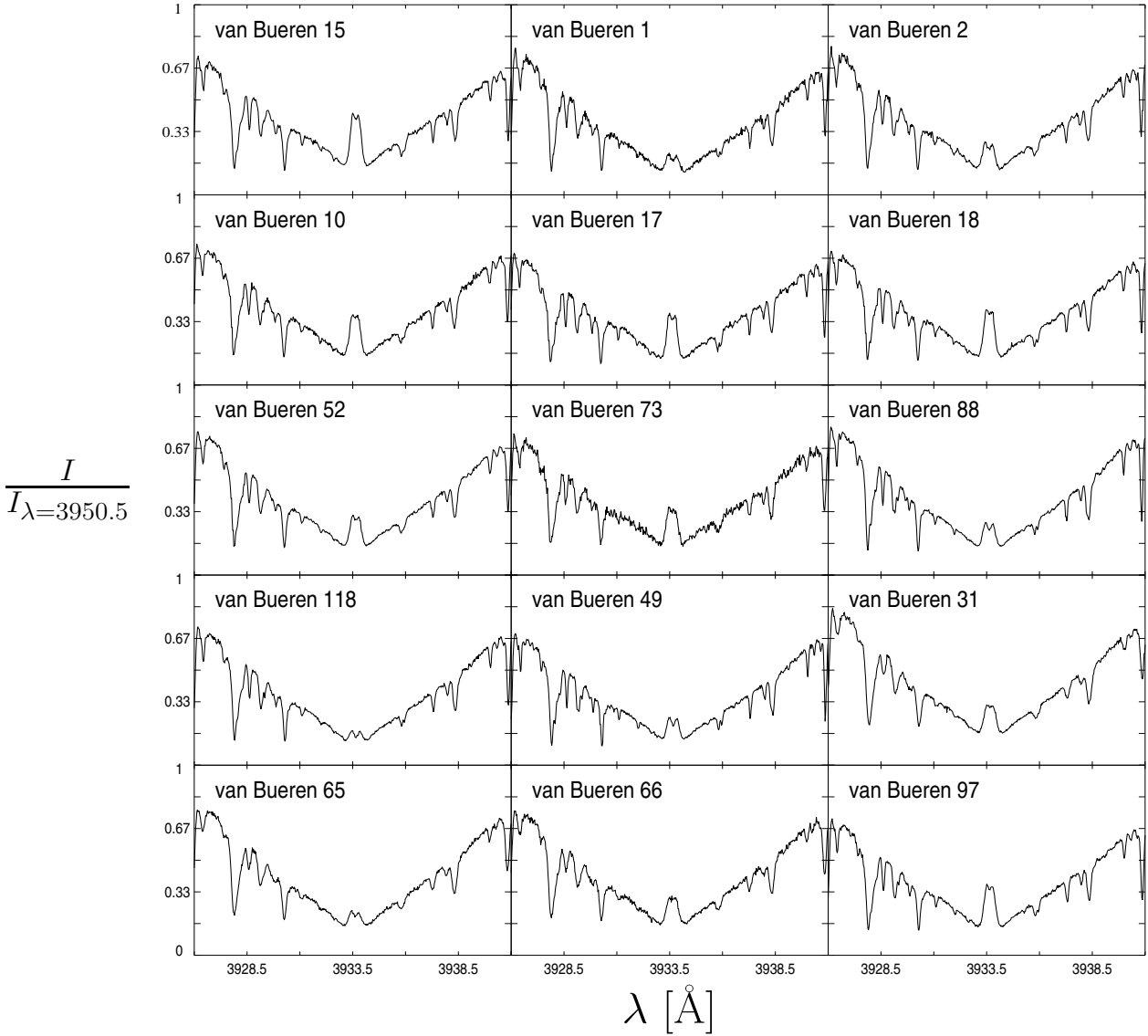}}\\
\\
{\huge $\lambda$  [\AA]}\\
\end{tabular}
\end{minipage}
}
\caption{HIRES spectra of the Hyades stars of our sample.}
\label{spectraHYADES}
\end{figure*}

\begin{figure*}
\begin{center}
\resizebox{16cm}{12cm}{
\begin{tabular}{c c}
&{\huge PRAESEPE}\\
&{\includegraphics[height=8cm]{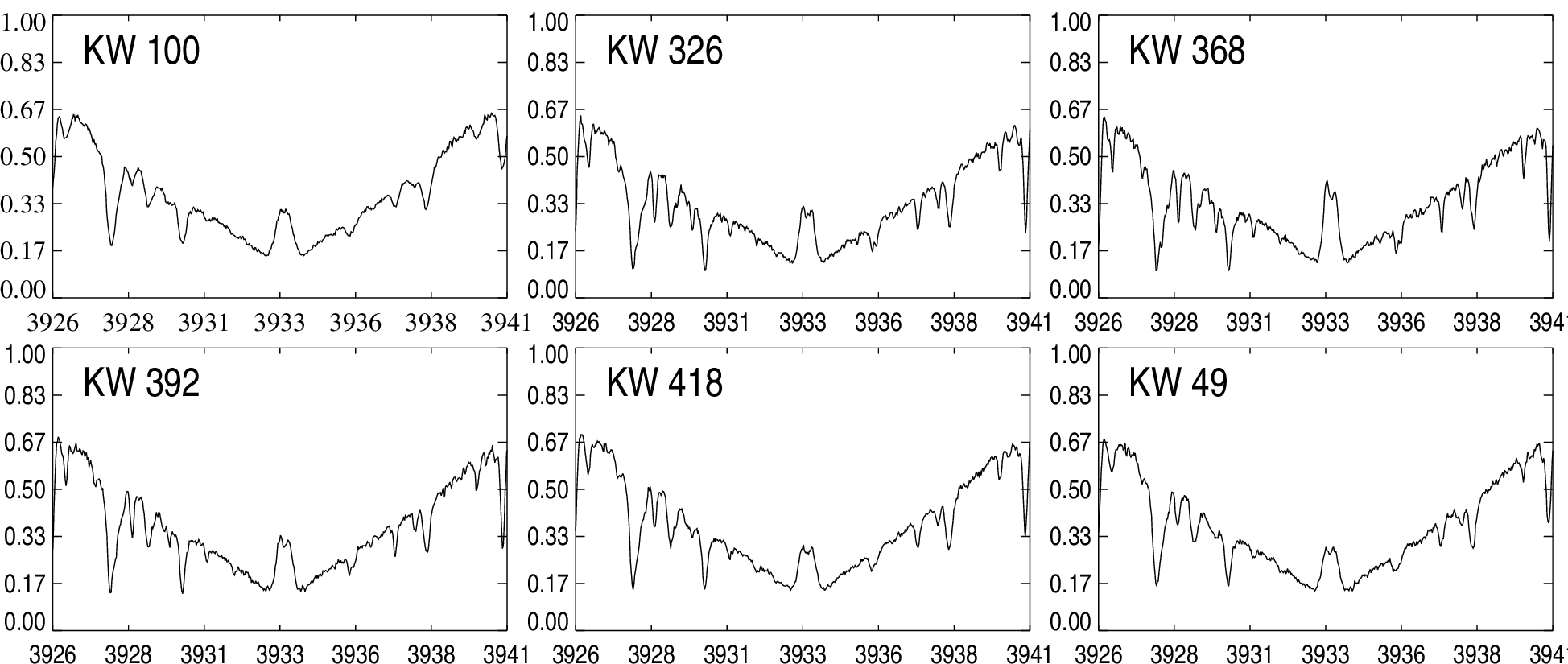}}\\
&\\
&\\
&{\huge IC 4651}\\ 
&\includegraphics[height=4cm]{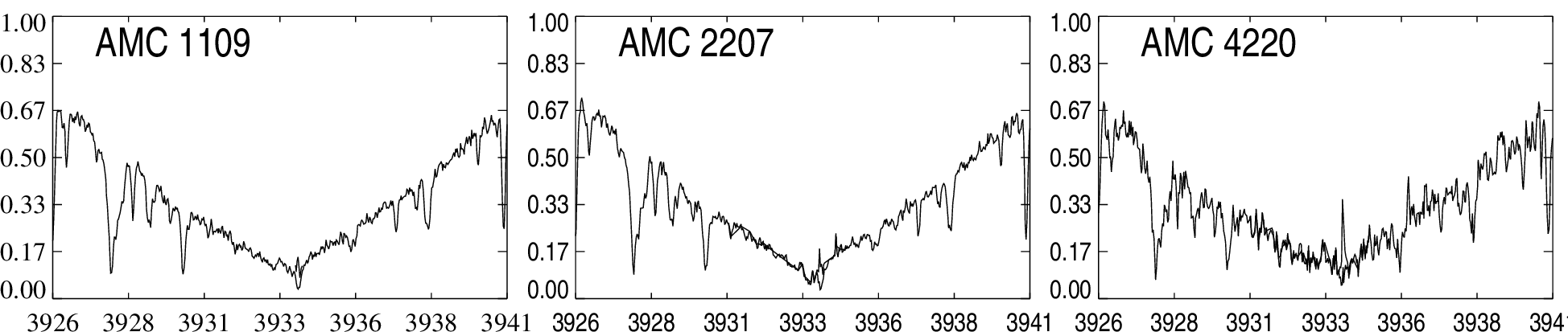}\\
&\includegraphics[height=4cm]{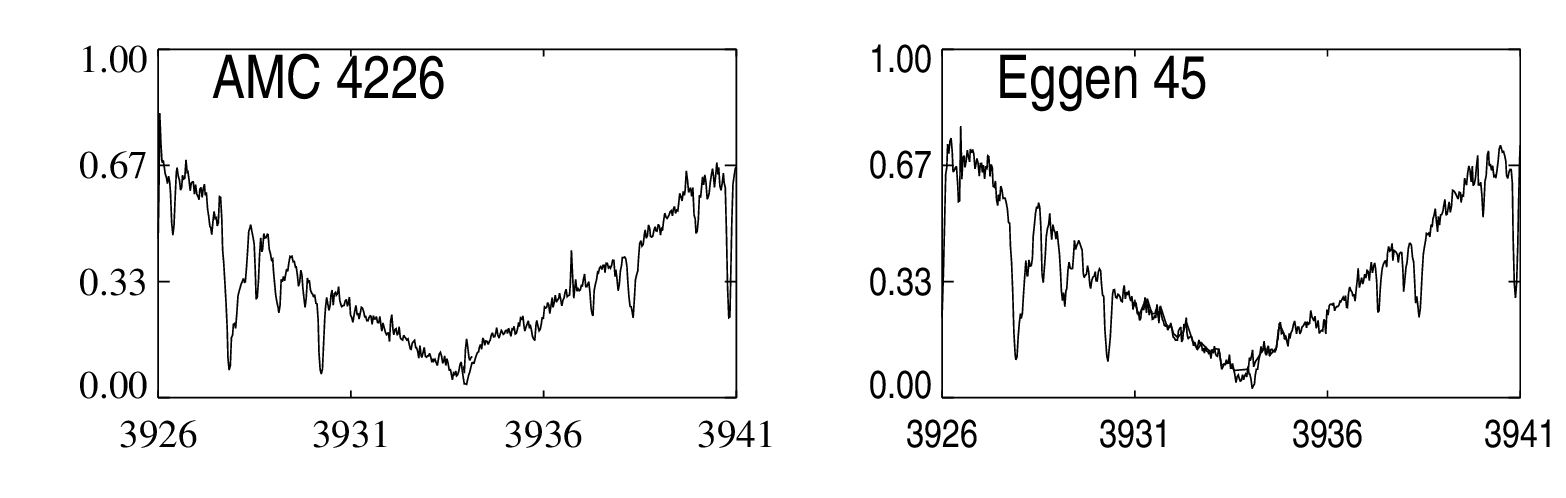}\\
&\\
&\\
{\huge $\frac{I}{I_{\lambda=3950.5}}$}&{\huge NGC 3680}\\
&\includegraphics[height=4cm]{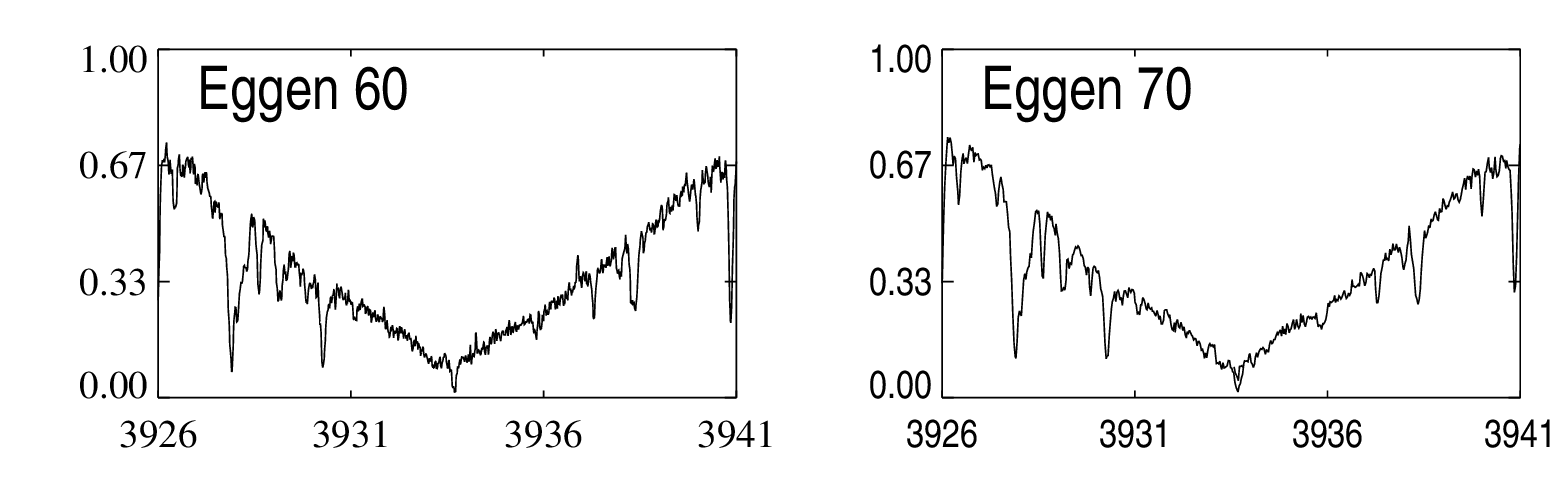}\\
&\\
&\\
&{\huge M 67}\\
&{\includegraphics[height=4cm]{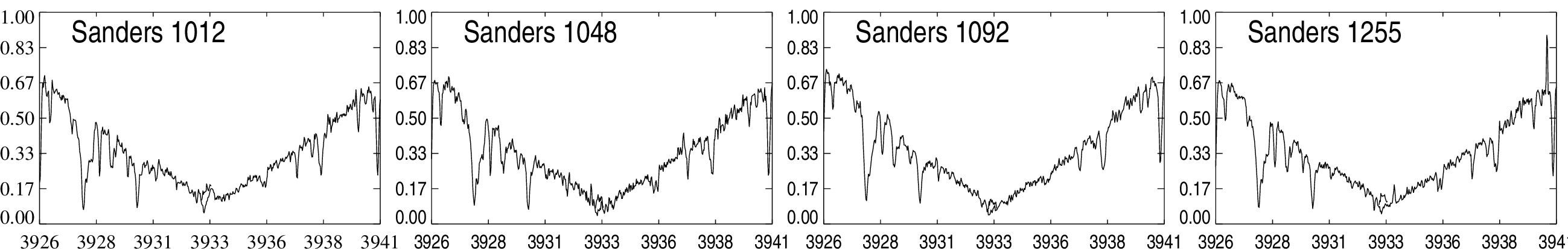}}\\
&{\includegraphics[height=4cm]{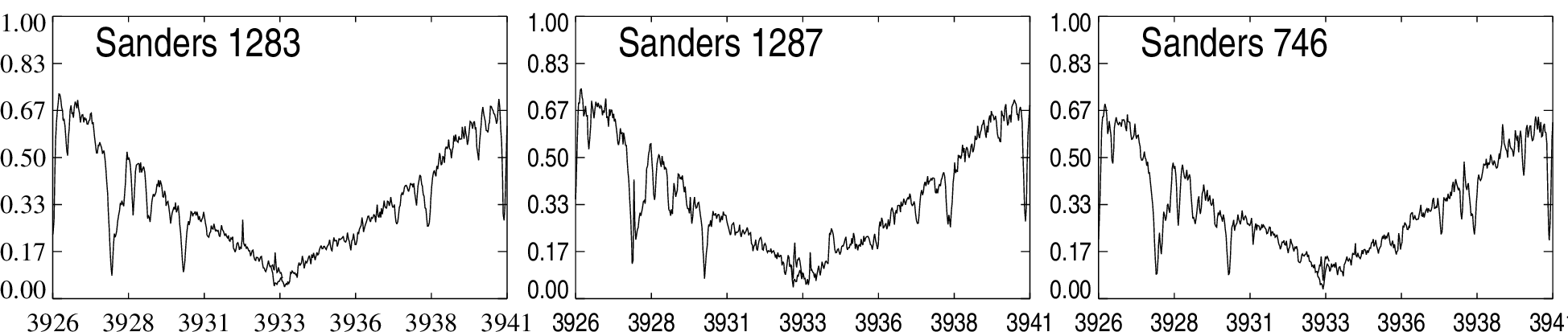}}\\
&\\
&\\
&{\huge $\lambda$  [\AA]}\\
\end{tabular}
}
\end{center}
\caption{UVES spectra of the stars of our sample. For the stars in IC 4651, NGC 3680 and M 67, in
each panel, the spectrum non corrected for IS line is plotted over the corrected one.}
\label{spectraUVES}
\end{figure*}

The observed spectra have been normalized to the 3950.5 {\AA} pseudocontinuum, as done in 
\citet{catalano78} and then in \citet{ppp88}. Since the spectra are the result of a merging of different echelle orders, 
we must check that merging is properly made. We did it in two ways: by observing 
early--type stars (i.e. stars with no Ca $\!{\rm II}$\  K stellar line), and by comparing the 
target spectra with 
a synthetic solar spectrum (courtesy of P. Bonifacio). This comparison of observed with photospheric--line 
profiles has been largely used by us, not only to verify that the rectification of the spectra 
was properly done, but also to check that other effects, such as 
background light, were correctly taken into account. The K line is very suitable 
for this purpose, because it is very deep, and therefore sensitive to small  mistakes 
in setting the zero flux level. 
In Figure \ref{avspectra} we show the average spectra of all the cluster stars and the solar one, 
all plotted over the synthetic solar Ca $\!{\rm II}$\  K stellar line. The solar spectrum is taken from the UVES archive
\footnote{{\small http:$//$www.eso.org$/$observing$/$dfo$/$quality$/$UVES$/$pipeline$/$}}.
We checked that the observed photospheric--line profiles successfully overlap after 
normalization. Of course this is only possible 
because all of the stars have similar colour to that of the Sun.
We have also checked the consistency of each observed K line profile with the synthetic solar K line profile. 
The spectra of the single stars are shown in Figures \ref{spectraHYADES} and \ref{spectraUVES}.

\begin{figure*}
\centering
\includegraphics[width=13cm]{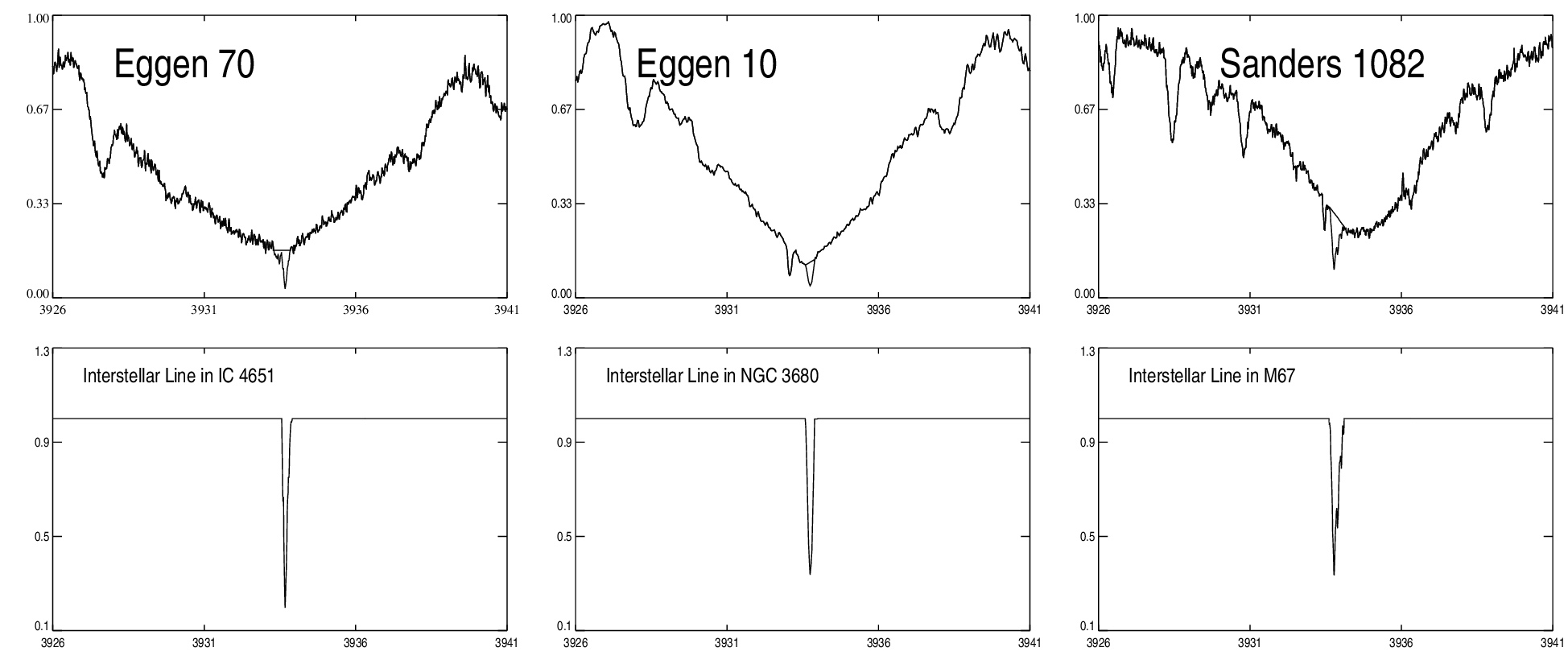}
\caption{Template spectra used to detect the IS--line feature.
The original spectra (one for each of the 3 distant clusters)
are over plotted on the  supposed non contaminated profile on the 3
upper panels. In the lower panels the resulting IS features
are shown.}
\label{islines}
\end{figure*}

For all clusters only one component of the IS Ca $\!{\rm II}$ K line falls in the stellar-spectra 
region where the 1--\AA\ index is measured.
In order to evaluate its contribution, we have observed for each of the
3 reddened clusters a reference star with an easily recognisable IS component. 
These stars show no narrow stellar line inside the Ca $\!{\rm II}$ absorption feature either because they are fast 
rotators (as in the case of Sanders 1082 in M 67, an Algool--type binary, and Eggen 70 in IC 4651) or because they are
hot enough (as in the case of Eggen 10 in NGC 3680) to present a pronounced broad deep core.
In Figure \ref{islines}, upper panels, the spectra of the quoted reference stars are shown. 
For each of them, the spectrum without the IS contribution is plotted over the original one. 
The latter has been divided by the former,
and the result is the IS contribution, which is shown in the bottom panel of Figure \ref{islines}.

Eggen 10 is quite hot, but this is not the case for Sanders 1082 and Eggen 70. If they were slightly active, their emission
would partially fill in the absorption IS line. In this case, the IS negative contribution would be
underestimated by the removal procedure. We are, however, confident that our estimate is quite robust, because Sanders 1082 and Eggen 70
are rather fast rotators, therefore if their chromospheric emission were relevant, they would be
as broad as the photospheric lines and the emission wings would be visible outside the IS contribution.

\begin{figure*}[ht]
\begin{center}
\resizebox{16cm}{4cm}{
\begin{tabular}{c c}
{\includegraphics[width=10cm]{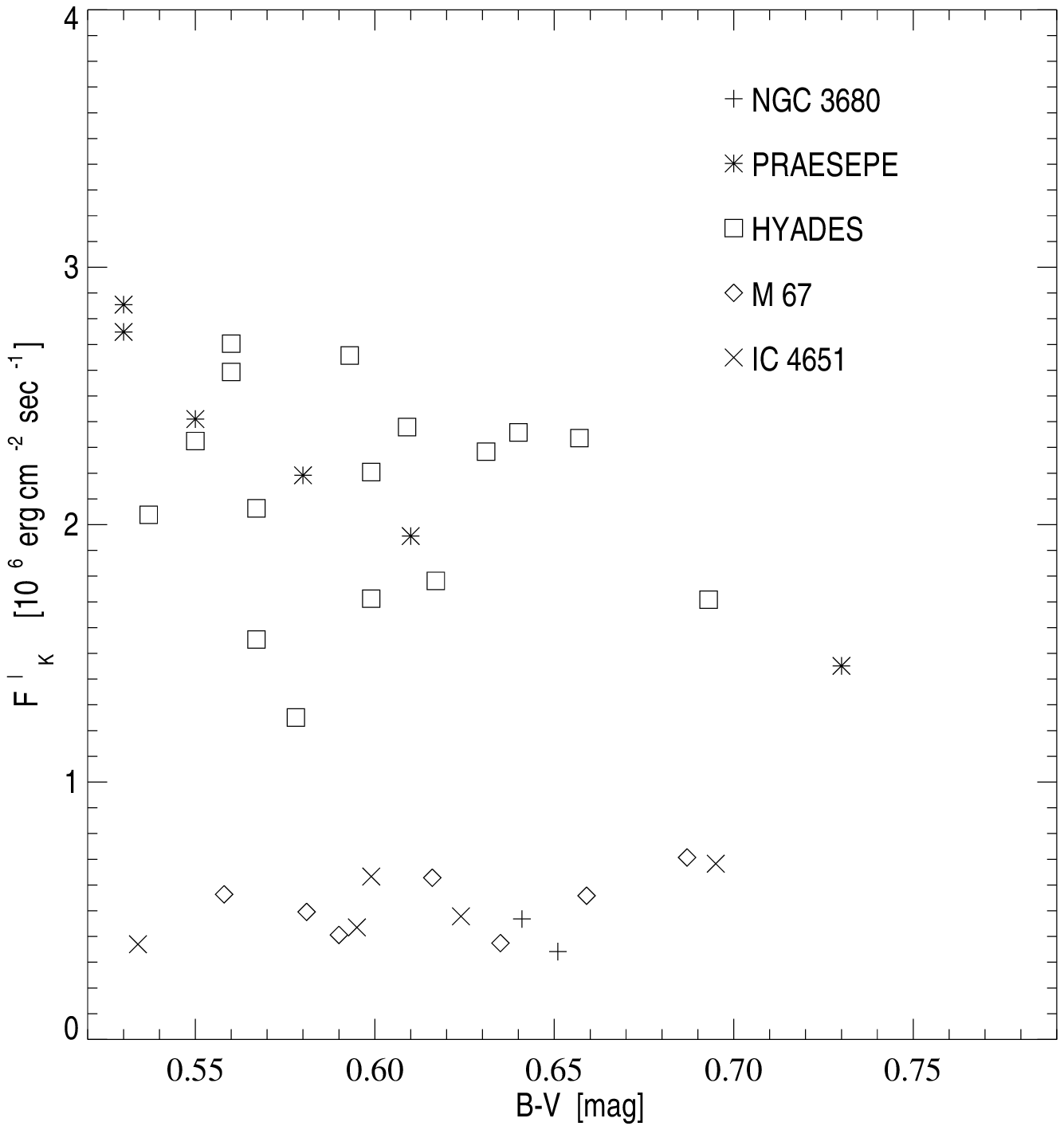}}&{\includegraphics[width=10cm]{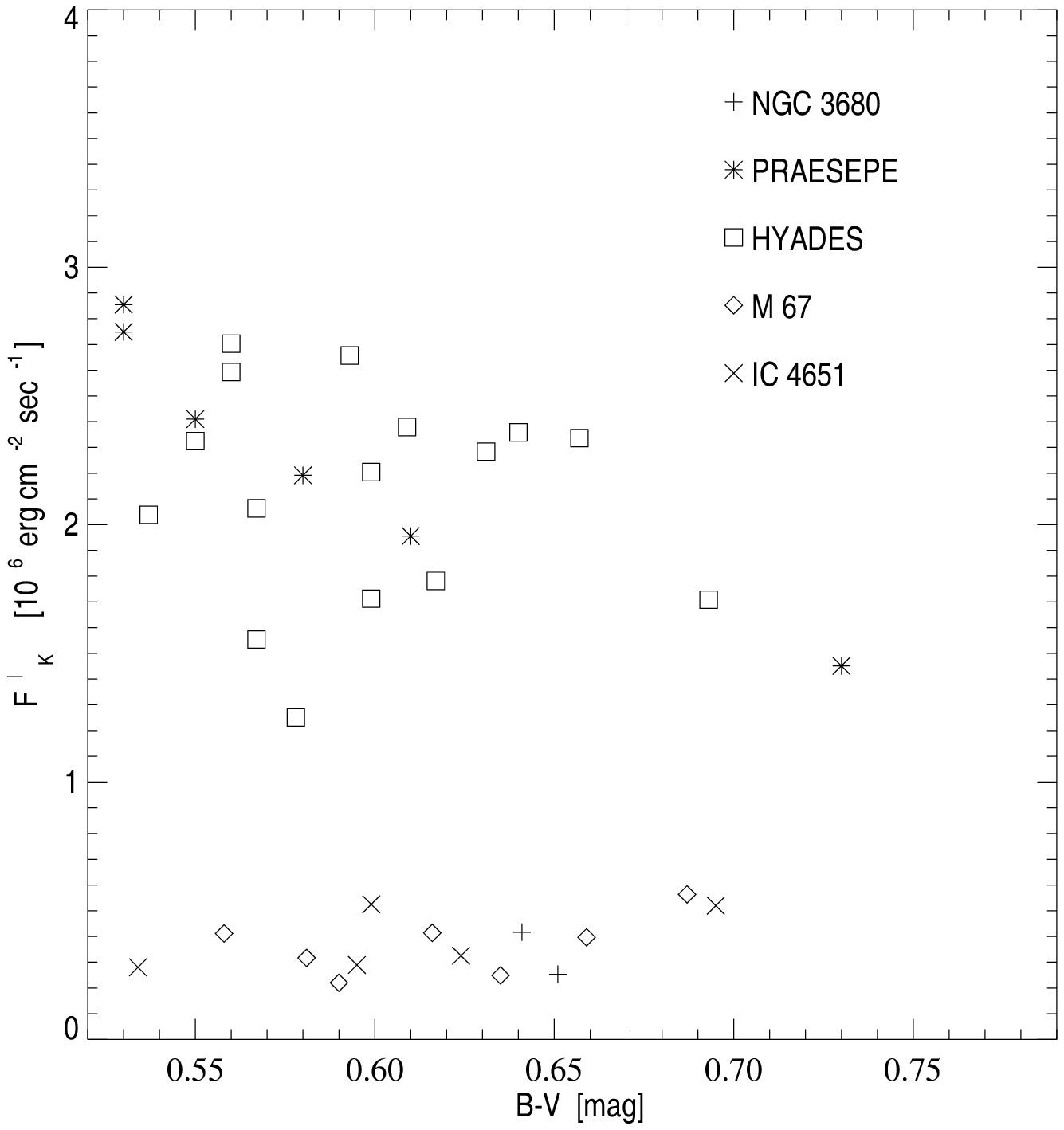}}\\
\end{tabular}}
\end{center}
\caption{The trend of $F_K^{\rm \prime}$ with the colour. In the right panel the 
chromospheric activity is not corrected for the 
IS absorption, in the left one chromospheric--activity indices regarding IC 4651, NGC 3680 and M 67 
stars are measured on  IS absorption corrected spectra. There is no remarkable difference between the two diagrams. 
A decreasing trend with colour for chromospheric--activity index can only be seen for Praesepe stars, 
since the range of colours  involved is narrow.} 
\label{BVvsCA}
\end{figure*}

Each K line spectrum of our target stars  has been divided by the appropriate IS spectrum, 
and the 1--\AA\  indices measured again. The results are given in the 
fourth column of Table \ref{maintab}. 
In Figure \ref{BVvsCA} we plot the  $F_K^{\rm \prime}$ vs.  B$-$V before and after the IS--line contribution 
has been subtracted (respectively left and right panels). Clearly the effect of this subtraction, although noticeable, 
does not affect  dramatically the relative emission between the different clusters, which is not surprising,
given the low reddening of the clusters.

Chromospheric fluxes decrease for cooler stars \citep[see e.g.][]{sdj91}.   
In Figure \ref{BVvsCA} such a trend is not present, with the possible exception of Praesepe stars.
By comparing our plot with that of \citet[][see Figure 6 therein]{sdj91}, we do not expect our results to be
significantly affected by the different colour distribution of the targets in the different clusters.

We have subtracted to the 1--\AA\ indices the photospheric contribution, 
calculated on the synthetic photospheric--solar model. The quantity subtracted amounts to 0.049 {\AA}.
Note that, even though this subtraction is the same for all of the stars, once the 1--\AA\ indices have been 
transformed into fluxes in the way explained later in this Section, the photospheric contribution becomes
colour dependent, because the transformation into flux is colour dependent.
The photospheric contribution we subtracted does not differ more than  20\%
from other approximations, e.g. the one used by \citet{nhbdv84}, and \citet{duncanetal}. 
The choice of another approximation does not significantly change the result,
even though it might affect the spread within each cluster.


We have  derived the chromospheric fluxes at the stellar surface transforming the 
IS corrected, photospheric subtracted equivalent widths into 
flux densities at the stellar surface (erg $\cdot$sec$^{-1}\cdot$ cm$^{-2}$).
This was made in three steps.
We first computed the V$-$R colours from the B$-$V ones.
We used the calibration in \citet{tesiluca}:
\begin{equation}
V-R=0.618 \cdot \left(B-V\right)^2-0.183\cdot \left(B-V\right)+0.375
\label{col_col_cal}
\end{equation}
We have  then used the calibration in \citet{ppp88} between
the V$-$R colour and the flux at the stellar surface of the pseudo--continuum radiation at 3950.5 
{\AA}:  
\begin{equation}
\log F_{3950.5}=8.459-2.833 \cdot \left( V-R\right)
\label{flux_col_cal}
\end{equation}
We have finally multiplied the equivalent widths for $F_{3950.5}$, obtaining the 
fluxes in the sixth Column of Table \ref{maintab}.

We need to estimate the absolute and relative accuracy of the indices and of the fluxes. 
Actually, the relative uncertainties are here the most relevant, since
we will ultimately compare the ratio of the fluxes between the different clusters. The absolute
value, although interesting, is just a normalised value for the purpose of the age--activity law. 

The Signal to Noise per pixel  is not the limiting factor of the accuracy, even though it is only of $\sim$ 10 in 
the spectra of the faintest stars, because we integrated on 1 {\AA}, namely $\sim$ 48 pixels and $\sim$ 37 
for HIRES spectra).
For the intermediate age clusters and M 67, the dominant sources of error in the 
measurements could come from an incorrect treatment of the inter--order (and scattered) light 
or from rectification problems. As we have mentioned before, we think we are rather free from 
these problems, because of the satisfactory match between the solar synthetic and stellar observed
profiles of the K line wings (i.e. the region between $\sim$ 3920 and $\sim$ 3950 {\AA}, outside the line core). 
We have first divided the photospheric spectra of all the stars of each cluster for their average spectrum, 
finding peak to peak differences of less than 7\% in their profile ratios. 
We have then repeated the same comparison with the   solar synthetic spectrum; and 
the differences between the stellar spectra and the solar model are greater. They arrive up to 
 20 \% for Hyades and Preasepe stars, while they are of up to about  10\% for the other clusters. 
Hyades and Praesepe
stars show exactly the same behaviour, despite the spectra were obtained with two different spectrographs.
The stellar spectra have always higher profiles than the 
solar model and that is true even for the Sun itself (cfr. Figure \ref{avspectra}).
This can be due to a low level of residual scattered light in the spectrograph, but the main reason is that 
the solar model we used does not include any chromospheric effect, which will tend to heat the higher layers 
of the photosphere and therefore to raise the photospheric profiles as well.
This effect is well known at least in the Sun, where the differences between the quiet and plage regions arise
also in the photospheric wings  of the K line \citep[][see Figure 1 therein]{labonte}.

Since the IS correction could be rather tricky, we have tried several different fits of the 
IS--line wings finding substantially only minor variations and we estimate the error from a wrong 
computation of the IS line as negligible. It is important to see also that this 
error is the same in both, absolute and relative terms and that the whole IS correction accounts for at most 
20$\%$ of the chromospheric flux measured.

The errors in $F_K^{\rm \prime}$  come from three main sources:
\begin{list}{-}{}
\item the uncertainty in the equivalent width measurement, which on its turn has, as mentioned, many other causes;
\item uncertainty in the V-R vs. B-V calibration;
\item the uncertainty in the B-V colours.
\end{list}

The absolute flux determination has an error of 30$\%$, due to the uncertainties in the  $\log F_{3950.5}$ vs.
$\left( V-R\right)$  calibration. 
However this affects mainly the zero point of our $F_K^{\rm \prime}$ measurements, and only to a much smaller extent
the relative  comparison.

The two other sources of error give a contribution to the relative uncertainty $\frac{\Delta F_K^{\rm \prime}}{F_K^{\rm \prime}}$
that can be calculated on the basis of Equations \ref{col_col_cal} and \ref{flux_col_cal}. They are, respectively:
$$\frac{\Delta EW}{EW}$$ and $$\left( 8.05 \cdot \left(B-V \right)-1.19\right)\cdot \Delta(B-V)$$

The typical relative maximum error in the equivalent width measurement of the single spectrum is within 10\%. For 
an error in the colour $\Delta(B-V)=0.04$, which we adopt as value of the maximum error in the colour, 
the corresponding relative error in the flux is, depending on B-V, between
the 12 and the 18\%. If we add quadratically
the two contributions, we finally obtain a maximum error in the single star measurement ranging from the 15 to the 20\%, 
if only relative values are concerned.

\subsection{Determination of the $v \sin i$ }
\label{rot}
As stated in Section \ref{data}, we have measured the FWHM of the cross correlation profile 
for all the spectra of our sample, and the data are shown in the seventh column of Table \ref{maintab}.

Line broadening, as extensively discussed e.g. in  \citet{gray}, 
is caused by several effects -- such as thermal motions, micro-- and macro--turbulence, Stark effect, 
Zeeman effect, natural broadening, instrumental response and rotational velocity.
We make two assumptions. The first is 
that the sum of these effects produces a gaussian shape, which is a reasonable 
assumption for solar--type stars \citep[see e.g.][]{ppp88}. The second is that the difference in broadening
among the stars of our sample is due only to  rotation, or  that the quoted effects 
contribute equally to the line broadening for all the stars, which is well justified given the 
similarity in the spectral type of the targets. 
We can now write:
$${\sigma_{CCP}}=\sqrt{{\sigma_{v \sin i}}^2+{\sigma_0}^2}$$ 
where ${\sigma_{CCP}}$ is the $sigma$ the cross correlation profile, provided that
it is a gaussian. For the reasons explained above, $\sigma_0$ is expected not to change significantly 
from star to star, within our sample.
Since ${\sigma_{CCP}}$ and $\sigma_{v \sin i}$ are proportional respectively to
the FWHM of the cross correlation profile and to ${v \sin i}$, we expect  $v \sin i$ to be:
\begin{equation}
v \sin i=A \cdot \sqrt{{FWHM}^2 - B^2}
\label{nonlincal}
\end{equation}
where $A$ and $B$, like $\sigma_0$, are constants. 
Using $v \sin i$ measurements from the literature we can calibrate
A and B in the previous relationship, and then calculate $v \sin i$ for all the 
stars. We have done this using the data of Table \ref{rotcaltab}, in particular the $v \sin i$ measurements 
in  the third Column and the FWHMs in the fourth Column. 

Most of the spectra of the calibrators have been obtained in a parallel project on IC 4651 \citep{przhcn}.
The spectrum of the Sun is from the UVES archive, and is the same plotted in the bottom panel
of Figure \ref{avspectra}. 
For Eggen 60 we used the FWHM of Table \ref{maintab},
since it is one of the targets. All the spectra used for the calibration are taken with the same instrument 
(UVES) at the same resolution (R=100000) of our observations and the same spectral range is used to get the cross 
correlation profiles.
The $v \sin i$ are taken from \citet{man02} for the stars in IC 4651 and from \citet{naa96} for Eggen 60.
For the Sun we adopted 2 Km/sec.

We decided not to use for the calibration Eggen 70 in NGC 3680, even though for this star  $v \sin i$ is 
available. The reason is that the 1 $\sigma$ error given for $v \sin i$ is three times 
larger than $v \sin i$ itself, that is 1 km/sec. 

The  calibration is a $\chi^2$ fit to the data of the Equation 
\ref{nonlincal}. 
The root mean square of the residuals of the data around the fitting curve is 0.26 km/sec, no star 
deviates more than  $10\%$ of its $v \sin i$, and 5 out of 8 calibration stars deviate less than $2\%$.

\begin{figure}
\centering
\includegraphics[width=8cm]{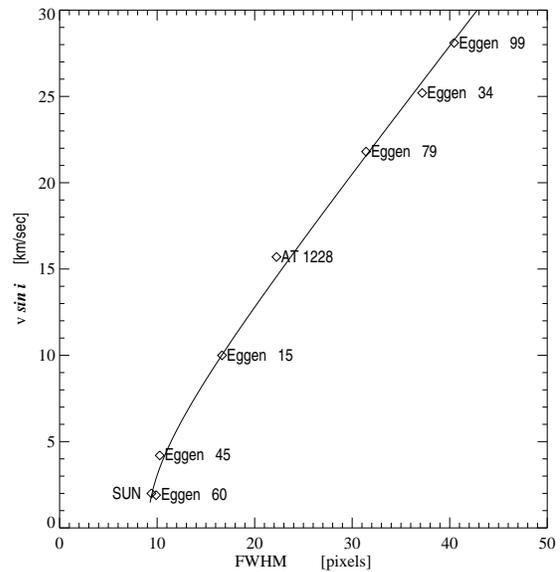}
\caption{Calibration of $v \sin i$ with the FWHM of the cross correlation profiles of UVES spectra.}
\label{rotcalfig}
\end{figure}

The result of the calibration is:
\begin{equation} 
v \sin i=0.72 \cdot\left( FWHM ^2- 9.05^2 \right) ^{1/2} 
\label{rotcaleq}
\end{equation}
The curve is plotted in Figure \ref{rotcalfig}.

We have used it to compute the values given in the eighth column of Table \ref{maintab}.
We remind the eventual users that we used the spectral range 588--680 nm, a spectral resolution
of R=100000, and a rebinning step of 0.0174 \AA

We note that the small residuals from our  $\chi^2$--fit possibly overestimates its real accuracy. 
The reason is that our calibration is a secondary calibration, because also the calibrating measurements 
are based on the FWHM  of the cross correlation profiles.
The errors may be quite larger in particular below $v \sin i \sim 3$ Km/sec, where other causes of line
broadening become more important. 
We underline that  we are mainly interested in differential values of $v \sin i$, that are as reliable as the  
 root mean square of our calibration indicates.

\begin{figure}
\centering
\includegraphics[width=8cm]{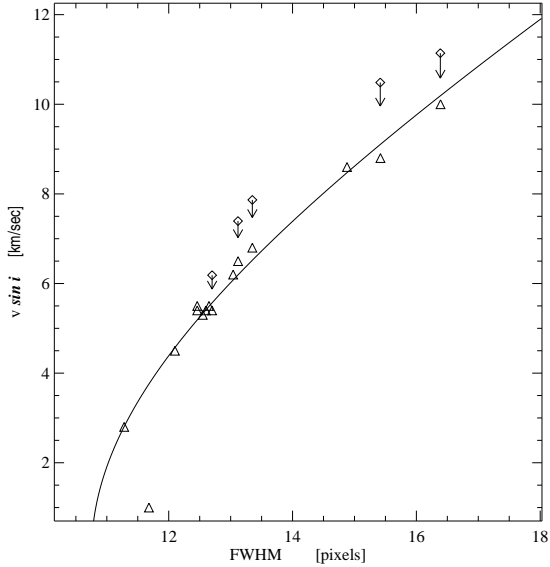}
\caption{$v \sin i$ vs. FWHM diagram for Hyades stars, and the fit to the 
data points: $v \sin i=0.82 \cdot\left( FWHM ^2- 10.75^2 \right) ^{1/2}$.
 Also equatorial velocities from periods of rotation
are plotted. The latter are indicated as square symbols with the downward arrow 
(they are in fact upper limits for $v \sin i$), while $v \sin i$ data points from \citet{psc03}
are indicated as triangles.}
\label{hyadesrot}
\end{figure}

Only for Hyades stars we could not use this calibration, since their spectra were taken with another 
spectrograph (HIRES) at a different resolution. Anyway, all of the $v \sin i$ for our Hyades targets are 
measured in \citet{psc03}. In this work, the rotational velocities are determined by comparing five synthetic 
Fe $\!{\rm I}$\  lines with the observed ones, therefore temperature is taken into account. 
In Figure \ref{hyadesrot} we plotted the diagram $v \sin i$ vs. FWHM. As for the calibration of UVES spectra,
we computed a fit of an equation of the type of Equation \ref{nonlincal} to the data points. The result is that
only one star deviates from the fit more than 5 \% of its $v \sin i$, and the median deviation is of the 2\%.
This proves that, for targets comprised in the short B$-$V colour range 0.51-0.73, the FWHM is virtually completely 
determined by $v \sin i$.
The presence of an outlier, namely van Bueren 88, confirms that more care has to be exercised in dealing with slow rotators.
We will not use the outlier value of $v \sin i$ in the analysis of the angular momentum evolution.
On the other hand, the overall quality of the data is evident when comparing Paulson et al.'s $v \sin i$ values 
with equatorial velocities from rotational periods.

\citet{rtldb87} provide precise period determination of 5 of our targets.
They used narrow band {\textit by} photometry to obtain light curves of 24 Hyades members. 
From the light curves it is possible to obtain the rotational period of the stars because rotational 
modulation of  photospheric features (spots) induces photometric variations. 
We transformed the periods into rotational velocities, and found the data in Table \ref{hyadesper}.
We had to estimate the stellar radii, 
which were computed using T$_{eff}$ and computed bolometric luminosity.
We transformed the B$-$V colours into  T$_{eff}$ and bolometric corrections using \citet{flower},
we then used Hipparcos parallaxes \citep{hipparcos} to transform the apparent magnitudes into luminosities.

\begin{table}[!h]
\begin{center}
\begin{tabular}{c c c c}
\hline
\hline
STAR     &  Cluster &$v \sin i$ {\tiny     [km/sec]} %
                           &  FWHM {\tiny [pixels]}\\
                                                    
\hline		     
AT  1228 & IC 4651  & 15.7 & 22.24\\ 
Eggen 15 & IC 4651  & 10.0 & 16.65\\
Eggen 34 & IC 4651  & 25.2 & 37.18\\ 
Eggen 45 & IC 4651  &  4.2 & 10.27\\
Eggen 79 & IC 4651  & 21.8 & 31.42\\
Eggen 99 & IC 4651  & 28.1 & 40.45\\
Eggen 60 & NGC 3680 &  1.9 &  9.89\\
Sun	 &  --      &  2.0 &  9.41\\

\hline
\end{tabular}
\end{center}
\caption{Literature data used for the calibration in Figure \ref{rotcalfig}. The range of B$-$V colours
is 0.514-0.641, closely encompassing the solar one.}
\label{rotcaltab}
\end{table}

\begin{table}[!h]
\begin{center}
\begin{tabular}{c c c c c}
\hline
\hline
van Bueren&Period         & inferred $v$   \\
id.       &{\tiny [days]} &{\tiny [km/sec]}\\
\hline 
31        &   5.45        & 11.1           \\
52        &   8.05        & 7.4            \\     
65        &   5.87        & 10.5           \\    
73        &   7.41        & 7.9            \\    
97        &   8.55        & 6.2            \\    
\hline
\end{tabular}
\end{center}
\caption{Rotational periods of  Hyades from \citet{rtldb87} and  rotational
velocities from them inferred. The data are 
collected to compare equatorial with projected rotational velocities.}
\label{hyadesper}
\end{table}

We overplotted the rotational velocities of Table \ref{hyadesper} on the diagram in Figure \ref{hyadesrot}.
They are equatorial velocities, i.e. upper limits of the projected 
rotational velocity $v sin i$, which is the quantity really related to the FWHM of the cross correlation profile. 
On the other hand, the fact that rotational modulation is observed indicates that the inclination angle of stars 
cannot be too small.
We therefore expect rotational velocities from period determinations to be right where they are: slightly above
the projected rotational velocities.

\section{Discussion}

After showing in the previous sections that we were able to derive chromospheric fluxes 
and rotational velocities with  great accuracy for our cluster stars, we can now
proceed to the discussion of the results. We first investigate the spread in the level of chromospheric 
activity in each cluster, which will tell to which extent a single 
value can be used for each cluster. By comparing the couple of twin points we can then evaluate if 
chromospheric activity is a function of age for solar stars. Only after that 
we will be able to derive an age--activity relationship. Finally, the same is applied to 
rotational velocities.

\subsection{Variability within each cluster}
\label{variability}
According to the detailed study of \citet{wl81}, the level of solar  chromospheric activity  varies of $\sim \pm 10\%$
(from peak to peak) during the 11--years solar cycle. If we consider the data corrected for photospheric 
contribution, then the variation becomes of $\sim \pm 20\%$. 
Analog cycles take also place in solar--type stars \citep[see e.g.][]{hss96,rlsb98,bdsh98}.
This poses some caution in deriving chromospheric ages from a single spectrum, and
causes scatter around the mean chromospheric--activity levels observed in each cluster, i.e. at a given age.
 
\citet{grhb00} have shown that indeed in the 
old M 67 considerable variations are observed if a sample large enough 
is acquired: they find that approximately 30$\%$ of the solar--type stars in M 67 exhibit 
levels of activity that are outside the envelope given by the solar--activity cycle.

\begin{table}[!h]
\begin{center}
\begin{tabular}{c c c}
\hline
\hline
cluster     &     spread &stars \\
            &
                          &      \\
\hline
  PRAESEPE  &     59.3\%   & 6    \\
    HYADES  &     68.3\%   &15    \\
   IC 4651  &     60.2\%   & 5    \\
  NGC 3680  &     31.2\%   & 2    \\
      M 67  &     62.4\%   & 7    \\
\hline	     
SUN         &     43.7\%   &      \\
\end{tabular}
\end{center}
\caption{Peak to peak spread of the $F_K^{\rm \prime}$ fluxes within each cluster, expressed
in percentage of the related mean value.
The spread of the solar data is the variation during the solar cycle 21.}
\label{spreads}
\end{table}

As shown in Table \ref{spreads}, all clusters but NGC 3680 (in which only two stars are observed) have a 
peak to peak spread slightly exceeding  $\pm 30\%$ of the mean chromospheric--activity level.
For young clusters such as the Hyades and Praesepe we expect that part of this variation is 
due to the superposition of long and short--term variability, the last induced by rotation. Another huge cause
of chromospheric variability in young stars is flaring activity.
For M 67, however, for which we expect the members to behave much like the Sun, the scatter is
1.4 times larger than the variations during a solar cycle.
This means that an additional spread of $\pm$ 24 \% has to be explained. For M 67, whose mean B-V colour is 0.62, 
the measurement maximum errors are of about the $\pm$ 18  \%.
This may indicate that
cycles similar to the 11--years solar cycle do not fully account for the dispersion of the data,
with interesting implications for the study of the Earth--Sun system.
Admittedly, 
we have observed each star only once (three consecutive times the fainter stars),
rather we assume that 
the stars in our sample have been observed at only one phase of their long--term activity, which
is likely different from star to star.
This implies, for instance, that possible star to star differences may cause additional noise
to our results. We think, however, that our procedure of overlapping the photospheric profiles
eliminates most of  the observational uncertainties. 
Before inferring firm conclusions, more data have to be collected.

\subsection{Hyades vs. Praesepe, IC 4651 vs. NGC 3680, M 67 vs. Sun}

\begin{table}
\begin{center}
\begin{tabular}{c c}
\hline
\hline
STAR           & $L_X$\\
&[$10^{28}$ erg/sec]\\

\hline
\hline
\multicolumn{2}{c}{PRAESEPE}\\
\hline

KW 49  & non det. \\
KW 100 &   6      \\
KW 326 & 3.20     \\
KW 368 & 34       \\
KW 392 & 10       \\
KW 418 & non det. \\
		
\hline		
\multicolumn{2}{c}{HYADES}\\
\hline		
van Bueren 1    &  3.7  \\  
van Bueren 2    & 15.1  \\   
van Bueren 10   &  5.8  \\   
van Bueren 15   & 12.9  \\
van Bueren 17   & 11.6  \\   
van Bueren 18   & 15.4  \\   
van Bueren 31   & 11.6  \\
van Bueren 49   &  7.0  \\   
van Bueren 52   & 10.8  \\   
van Bueren 65   & 10.7  \\
van Bueren 66   & 17.1  \\
van Bueren 73   & 11.2  \\   
van Bueren 88   &  --   \\   
van Bueren 97   &  7.7  \\
van Bueren 118  &  5.3  \\   

 \hline
\end{tabular}
\end{center}
\caption{X--ray flux measurements of Praesepe and Hyades stars. All of them but one refer to the ROSAT detection
 in the band 0.1--2.0 keV. KW 326 was not detected by ROSAT. Its measurement, which refers to the same band, has been 
performed with XMM--{\textit Newton} satellite.}
\label{xraydata}
\end{table}

\begin{figure*}[ht]
\centering
\includegraphics[width=15cm]{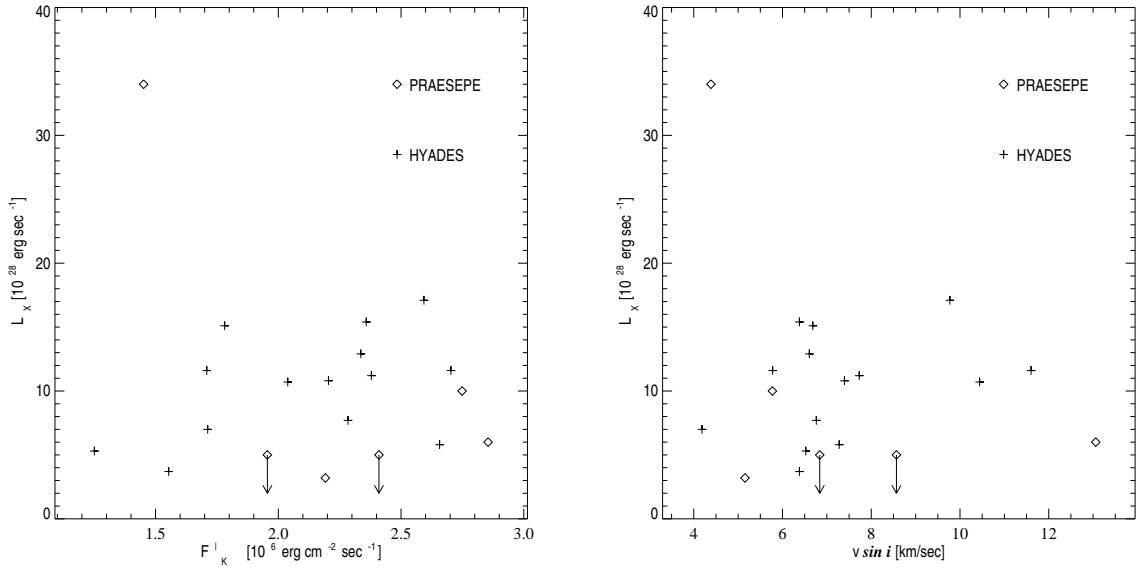}
\caption{X--Ray luminosities vs. chromospheric activity and rotation diagrams.
 In neither of those a trend can be clearly seen.}
\label{CAvsXRAY}
\end{figure*}

The Hyades and Praesepe clusters are particularly relevant because they
have been studied also in the X--ray \citep{rs95}. 
It is at least reasonable to reckon that the X--ray emission (i.e. coronal activity) and the chromospheric activity
are not completely independent, since they are both related to rotation and  magnetic field 
\citep[see e.g.][and references therein]{bks01}, but in 
\citet{rs95} the two coeval clusters  were found to have substantially different levels of X--ray emission \citep{rs95}. 
This early result has been recently revised by \citet{frp03},
who find that Praesepe stars have the same level of X--ray activity of Hyades if the central field is considered.

Our data confirm that Praesepe and Hyades stars span the same ranges of chromospheric--activity levels 
and $v \sin i$ values, even though our Praesepe stars have a mean X--ray luminosity significantly lower than 
the Hyades ones (less than a half).

In Figure \ref{CAvsXRAY} the $L_X$ vs. $v \sin i$ (right panel) and the $L_X$ vs. $F_K^{\rm \prime}$ diagram (left panel) 
are shown for both clusters.  $L_X$ measurements are taken from from \citet{ssk95} for the Hyades, and \citet{rs95} for 
Praesepe members, but KW 326, which is given in \citet{frp03}

In both panels the spread is high and no real relationship is observed. 
In spite of the aforementioned possible discrepancy between X--ray emission and chromospheric activity, a significant
correlation likely exists. The fact that it is not seen in Figure \ref{CAvsXRAY} could be due to the  
activity  variability, since the optical and X--ray observations are not simultaneous.

When focusing on the other coeval clusters we notice in Table \ref{maintab} that the activity and rotation values 
for IC 4651 and NGC 3680 are, as in the  in the case of Praesepe and Hyades, indistinguishable. 
We have to remind the reader that only two stars were observed
in NGC 3680, therefore its average chromospheric--activity and rotation levels cannot be considered accurate. 
Nevertheless it is very significant that both  NGC 3680 stars have values
similar to those of IC 4651.
This result is particularly important because, as we shall 
see in the following, these clusters show a novel behaviour with respect of what thought so far.

Finally, given the excellent  agreement between the mean M 67 activity values and the Sun, we can conclude
that all our coeval points show very consistent average values, therefore the level of chromospheric activity is an
age driven parameter.

The $v \sin i$ for M 67 is slightly larger than the solar one ($\sim$ 3 km/sec, see Section \ref{rot}); we do not consider
this at the moment  as a significant discrepancy, but it would be interesting to verify that  M 67 cluster stars rotates 
indeed slightly faster than  the Sun.

\subsection{The age--activity--rotation relationship.}
\begin{figure*}[ht]
\centering
\includegraphics[width=13cm]{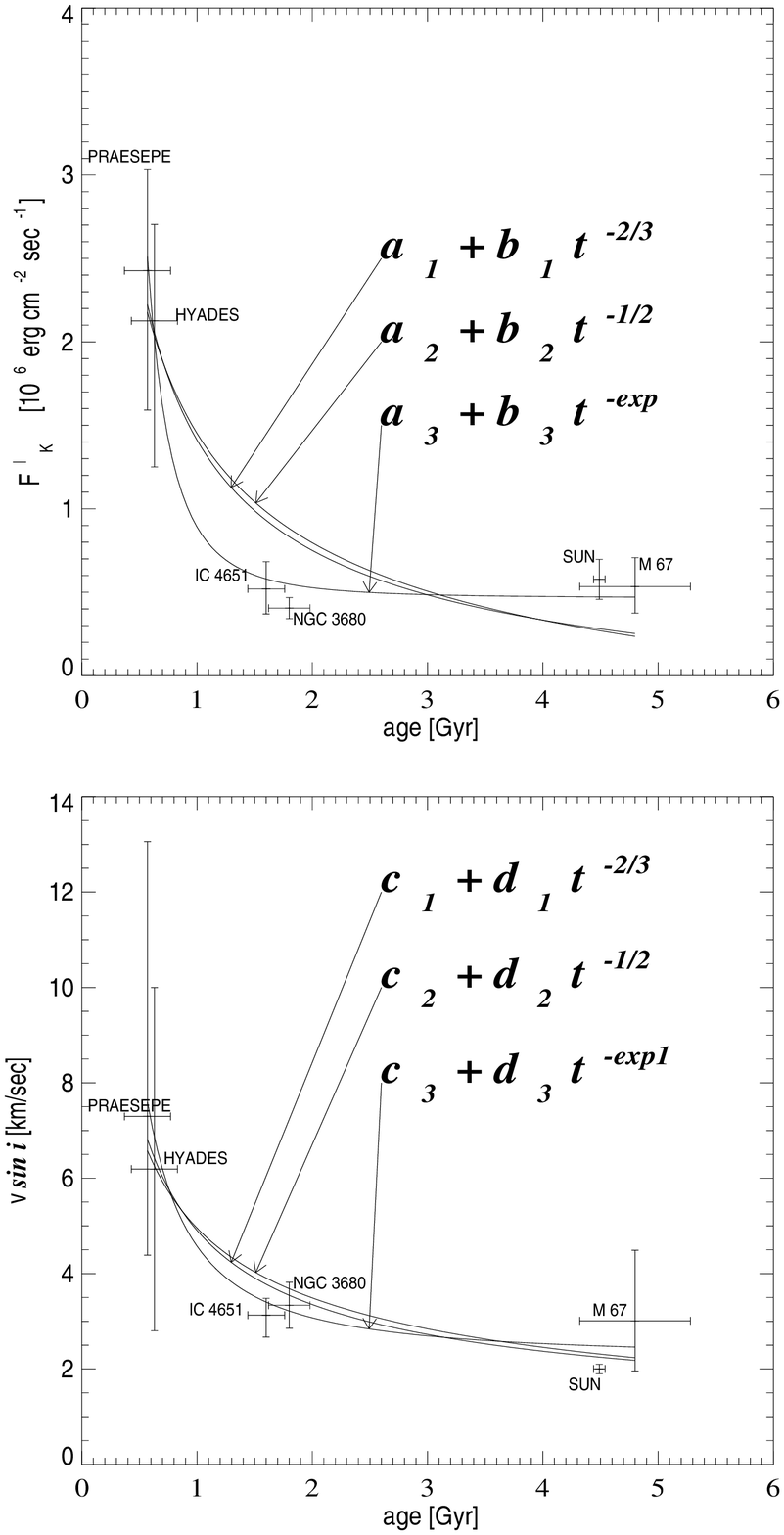}
\caption{The evolution of chromospheric activity (measured by $F_K^{\rm \prime}$
corrected for IS absorption, 
upper panel) and rotation (lower panel) with age. Here 
the vertical bars represent
the whole spread of the ordinate data within the cluster. The points are the mean values. For both quantities the 
surprising result is that the intermediate age clusters show the same level of chromospheric activity and 
mean rotational velocity of M 67 and the Sun. 
Since, as remarked, the bars are not error bars but peak to peak spreads, their position attests that
the activity distribution of the 7 intermediate age stars and that of the 7 M 67 stars are virtually equal.
For both diagrams we tried to fit three power laws, as indicated in the figure. -2/3 and -1/2 are the exponent values
suggested respectively by \citet{bch87} and \citet{skumanich}. As can be seen, they do not fit our data.
By minimizing the $\chi^2$ we find, for the free parameters: $a_1=-0.38, b_1=1.79, a_2=-0.79, b_2=2.24, 
a_3=0.47, b_3=0.43, exp=2.79, c_1=0.71, d_1=4.20, c_2=-0.05, d_2=5.00, c_3=2.23, d_3=2.35, exp1=1.47$.}
\label{CAvsAGE}
\end{figure*}

When the stars of about 1 M$_{\odot}$ reach the ZAMS, they show a high dispersion  in rotational velocity, and therefore in
chromospheric--activity levels, as a result of their different Pre Main Sequence histories  \citep[see e.g.][]{bouvier}. 
But at 600 Myr, i.e. 
the youngest age we are dealing with in the present study, the magnetic braking mechanism already made the rotational velocity uniform.
So, for each of our clusters, rotational velocity and chromospheric activity are expected to be all of 
the same order of magnitude (see Section \ref{variability}).
The evolution of these quantities, as it results from our data, is shown respectively in the two panels 
of Figure \ref{CAvsAGE}. 
It is very important to bear in mind that in this Figure the bars are not error bars, but they represent the 
peak to peak spread within the clusters, 
i.e., the difference between the maximum and the minimum  cluster values, and the points are the mean values.

As expected, chromospheric activity and rotation show the same decaying trend. 
Surprisingly, however,  the intermediate age clusters IC 4651 and NGC 3680 have the same
level of activity and rotation as  the older M 67 and the Sun. 
We recall that this two data points are based on
measurements of 7 stars, two in NGC 3680 and 5 in IC 4651, none of which shows a chromospheric 
activity above the most active star in M 67, that is to say that the activity distribution of the
7 intermediate age stars and that of the 7 M 67 stars are virtually equal. 

This means that an abrupt decay of chromospheric activity occurs between Hyades' age and 1.5 Gyr, after which 
a very slow decline follows, if any. 
Three power law decay laws are plotted over the data points of Figure \ref{CAvsAGE}: 
the two more gentle slopes represent the 
$t^{-1/2}$  and $t^{-2/3}$ laws proposed respectively by  \citet{skumanich} and \citet{sdj91}, scaled 
to best match our data points. 

The intermediate age clusters are  clearly not consistent with any of the laws so far suggested; they have
an activity level lower more than a factor 2.
If the exponent of the power law is left as a free parameter, we find that the points are best 
fitted by the third curve 
plotted on the $F_K^{\rm \prime}$ vs. age diagram, namely $F_K=0.5\cdot t^{-2.5}+0.5$, where $F_K^{\rm \prime}$ is in unit of 
10$^6\cdot$erg$\cdot$cm$^{-2}\cdot$sec$^{-1}$, and the age is expressed in Gyr.
Note that neither a different zero point, nor different assumptions
on the age of the clusters would change our conclusion, since activity keeps constant
for a time interval, the one between the intermediate age clusters and M 67, that under no acceptable assumptions 
can be shorter than 2 Gyr. Similar results are also claimed by \citet{rlp03}.

Figure \ref{CAvsAGE} recalls what observed among field stars: the so called Vaughan--Preston gap, 
i.e. the lack of nearby stars with a level of chromospheric activity intermediate
between Hyades'  and the Sun \citep{vp80}. With our results this gap gets a quite natural explanation: 
the decline in activity is so abrupt to be almost a discontinuity, and 
only stars with ages  in the range between 0.6--1.5 Gyr may fall in it. 

As mentioned above, the previous most systematic attempt to study chromospheric activity in stars belonging to open clusters 
has been made by \citet{bch87}, who obtained results quite different from ours. 
Some major discrepancy may exist in the measurement of the activity index. For instance,
in \citet{bch87},  the ratio between the  chromospheric fluxes 
in the Hyades  and in M 67 stars or in the Sun, is only $\sim 2.2$, while for us is larger than 4.
We find a similar ratio ($\simeq$ 2.4) if our index includes the photospheric
contribution which represent a fraction of the 1--\AA\ index much larger for M 67 than for Hyades stars.
In Figure \ref{CAvsAGE} the difference between ours and Barry et al.'s data is apparent.
Given the large scatter of fluxes observed in each cluster, we cannot, however, exclude that the discrepancy is due to 
differences in the sample observed. We tend to consider this  possibility unlikely, considering the large number of stars observed.
Barry et al. have observed 13 stars in the Hyades and 16 in M 67 and we have observed 15 stars in Hyades and 7 in M 67.
These authors measure also a chromospheric--activity index for the intermediate age cluster  
NGC 752 only 1.4 times lower than  the Hyades and  1.5 times bigger than  M 67; this would imply 
a level of chromospheric activity in the middle between the Hyades and the old stars
plateau, right in  the Vaughan--Preston gap.
\citet{cc94}, so far used as age reference, evaluate the age of NGC 752  comparable, but 
 slightly younger than IC 4651 (1.5 vs. 1.6 Gyr).

Before going into speculations about the difference between our and
Barry et al.'s results, the NGC 752 chromospheric--activity level 
should be proven by means of high--resolution spectroscopy in solar analogs.

\subsection{The impact of our results on the determination of ages and on the age of galactic population}
\begin{figure*}[ht]
\centering
\includegraphics[width=16cm]{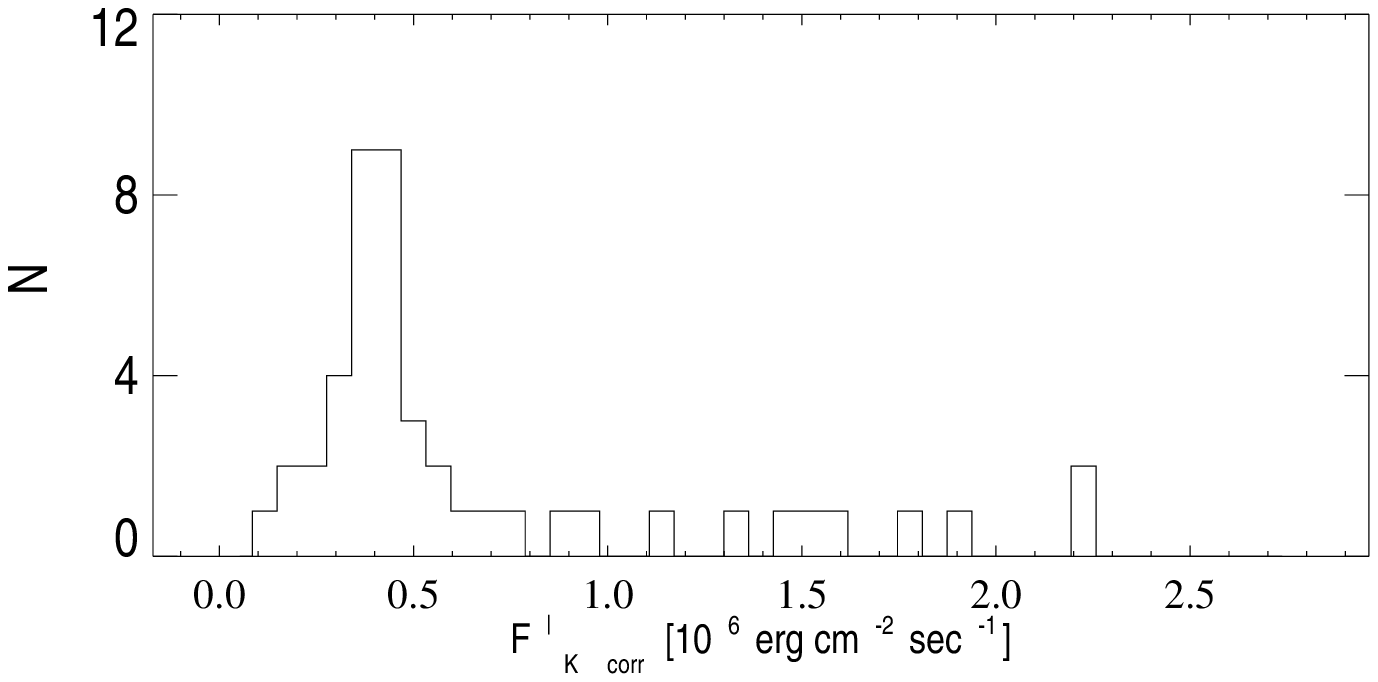}
\caption{Histogram of the chromospheric--activity levels in the sample of stars from \citet{luca92}.}
\label{istoluca}
\end{figure*}

According to the picture we have drawn in the last Section, 
stars can be divided in three groups
on the basis of their chromospheric--activity levels:
\begin{list}{-}{}
\item the youngest stars, with $F_K^{\rm \prime} \gtrsim 1.2 \cdot 10^6$ erg  cm$^{-2}$ sec$^{-1}$
\item the stars with  $F_K^{\rm \prime}$ between $\sim$ 0.7 and $\sim$1.2 erg  cm$^{-2}$ sec$^{-1}$
\item the oldest stars, with $F_K^{\rm \prime}   \lesssim 0.7 \cdot 10^6$ erg  cm$^{-2}$ sec$^{-1}$
\end{list}

Among our target stars, those in Praesepe and Hyades belong the first group, all the others to the third one.
Let us call the upper age limit of the first ensemble $t_1$ and the lower age limit of second $t_2$.
We don't know the exact values of $t_1$ and $t_2$. We only know that 0.6 Gyr $< t_1 < t_2 < 1.6$ Gyr.

In Figure \ref{istoluca} we plot the histogram for 44 solar--type stars (spectral type between G0 and G5) in the solar neighborhood taken
from \citet{luca92}.
The numbers of stars belonging to each group are respectively: 8, 2 and 34. The result confirms what expected: most of the stars 
have an age older than  $t_2$, much less are younger than $t_1$, only $\sim 5\%$ have ages between $t_1$ and $t_2$, i.e. those in the 
Vaughan--Preston gap.
In addition, the standard deviation of the $F_K^{\rm \prime}$ values of the 34 less active stars in 
the sample of \citet{luca92} is very similar to that of the 14 targets belonging to the three intermediate age and old clusters 
IC 4651, NGC 3680 and M 67: 0.12 $10^6$ erg  cm$^{-2}$ sec$^{-1}$.
Unfortunately our results indicate that an accurate age determination based on chromospheric activity may not be obtained for stars
older than $\sim$ 1.5 Gyr. 
Once $t_1$ and $t_2$ are determined precisely, much could be said
about the recent Star Formation History in the solar neighborhood, provided that initial conditions do not dominate the
fast decline of chromospheric activity occurring between $t_1$ and $t_2$.


\section{Conclusions}
We have studied the evolution of chromospheric activity and rotation in solar--type stars by analysing 
VLT and KECK high--resolution 
spectra of  35 stars belonging to 5 open clusters and published solar data. 
As a measure of chromospheric activity we have used $F_K^{\rm \prime}$, i.e. 
the equivalent width of the Ca $\!{\rm II}$ K line emission core corrected for photospheric
contribution and transformed into flux at the stellar surface.  
Our data set can be divided in three age bins: 6 Praesepe stars and 15 Hyades stars at 0.6 Gyr;
5 IC 4651 stars and 2 stars in NGC 3680, at about 1.7 Gyr;  and finally 7 M 67 stars and the Sun, at about 4.5 Gyr.
We undertook the present research mainly to ascertain a decay law which allows measuring chromospheric 
ages in field stars.
We have found that stars within each age bin  show indeed the same mean level of chromospheric activity, confirming
that it is possible to search for a  chromospheric activity--age relationship. 
Stars in the youngest bin have chromospheric fluxes
ranging between 1.2 and 2.8 10$^6$ erg  cm$^{-2}$ sec$^{-1}$. All the other stars have values ranging 
between  0.3 and 0.7 10$^6$  erg  cm$^{-2}$ sec$^{-1}$. The most important result is that there is no difference in the distribution of
chromospheric--activity levels between the intermediate and the oldest age bin, even though  difference of ages
of the two bins is as large as $\sim$ 3 Gyr. 
We obtain  a very similar conclusion about the angular momentum evolution, with the only difference
that the gap between the youngest bin  and the others is less pronounced. 
A fast decay of  chromospheric activity occurs roughly between 0.6 and  1.5 Gyr of
Main Sequence lifetime, after which a kind of plateau appears. 

No power law so far proposed can fit our data. The best match curve we computed, namely 
$F_K^{\rm \prime}=0.5\cdot t^{-2.4}+0.4$,  where $F_K^{\rm \prime}$ is in unit of 
10$^6\cdot$erg$\cdot$cm$^{-2}\cdot$sec$^{-1}$, and the age is expressed in Gyr, can reproduce this trend.

Coupling this trend with the observational fact that in each of the clusters observed 
the star to star activity level vary almost by  a factor 2,  our results indicate that 
the determination of chromospheric ages for field stars older than $\sim$ 1.5 Gyr
is extremely challenging, or even 
potentially misleading in particular if only one or a few activity measurements are available.

\begin{acknowledgements}
We are greatly indebted to Bill Cochran, Artie Hatzes and Diane Paulson, for making available  the HIRES spectra of Hyades' stars,
to Piercarlo Bonifacio, for computing the  synthetic Ca $\!{\rm II}$\  solar spectrum, and to Claudio Melo,
for sharing the  code and the digital mask for the computation of  the cross correlation profiles.
This paper took advantage of the useful comments of Diane Paulson and of the impressive work of the referee, Dr. Messina.
We also thank Nick Pettefar for the careful reading of the manuscript.
\end{acknowledgements}

\bibliographystyle{aa}

\begin{thebibliography}{}

\bibitem[Allen, 1973]{allen}
  Allen, C.W., 1973, Astrophysical Quantities (The Athlon Press)

\bibitem[Anthony Twarog \& Twarog, 1987]{AMC1}
  Anthony Twarog, B., \& Twarog, B. A. 1987, AJ 94, 1222 

\bibitem[Anthony Twarog et al., 1988]{AMC}
  Anthony Twarog, B., Mukherjee, K., Caldwell, N., \& Twarog, B. A. 1988, AJ 95, 1453 

\bibitem[Ballester et al., 2000]{bmbchkw00}
  Ballester, P., Modigliani, A., Boitquin, O., Cristiani, S., Hanuschik, R.,
  Kaufer, A., Wolf, S., 2000 ESO Mess. 101, 31

\bibitem[Baliunas et al., 1998]{bdsh98}
  Baliunas, S.L., Donahue, R.A., Soon, W., \& Henry, G.H. 1998, ASP Conference Series 154, 153 

\bibitem[Barrado y Navascues et al., 1997]{bsr97}
  Barrado y Navascues, D., Stauffer, J.R., \& Randich, S. 1997, MmSAI 68, 985

\bibitem[Barry et al., 1987]{bch87} 
  Barry, D.C., Cromwell, R. H., \& Hege, E. K. 1987, ApJ 315, 264

\bibitem[Benz \& Mayor, 1981]{bm81}
  Benz, W., \& Mayor, M. 1981, A\&A 93, 235

\bibitem[Blanco et al., 1974]{bcmr74}
  Blanco, Catalano, Marilli, \& Rodon\'o 1974, A\&A 33, 257

\bibitem[Bouvier, 1997]{bouvier}
  Bouvier, J. 1997, Mem.S.A.It. 68, 881

\bibitem[Bruevich et al., 2001]{bks01}
  Bruevich, A.E., Katsova, M.M., \& Sokolov, D.D. 2001,  Astronomy Reports 45, 718

\bibitem[Carraro \& Chiosi, 1994]{cc94} 
  Carraro, G., \& Chiosi, C. 1994, A\&A  287, 761

\bibitem[Carraro et al., 1996]{cgbc96} 
  Carraro, G., Girardi, L., Bressan, A., \& Chiosi, C. 1996, A\&A  305, 849

\bibitem[Catalano, 1978]{catalano78}
  Catalano, S. 1978, A\&A 80, 317

\bibitem[Cochran et al., 2002]{chp02}
  Cochran, William D.; Hatzes, A. P.; Paulson, Diane B. 2002, AJ 124, 565

\bibitem[Dekker et al., 2000]{ddkdk00}
  Dekker, H., D'Odorico, S., Kaufer, A., Delabre, B., \&  Kotzlowski, H. 2000, SPIE 4008, 534 

\bibitem[Duncan et al., 1984]{duncanetal}
  Duncan, D.K., Baliunas, S.L., Noyes, R.W., Vaughan, A.H., Frazer, J., \& Lanning, H.H. 1984, PASP 96, 707

\bibitem[Dupree et al., 1999]{dwp99}
  Dupree, A.K., Whitney, B.A., \& Pasquini, L. 1999, ApJ 520, 751

\bibitem[Eggen, 1969]{eggen69}
  Eggen, O.J. 1969, ApJ 155, 439

\bibitem[Eggen, 1971]{eggen71}
  Eggen, O.J. 1971, ApJ 166, 87

\bibitem[ESA, 1997]{hipparcos}
  ESA, 1997, The Hipparcos and Tycho Catalogues, ESA SP--1200

\bibitem[Flower, 1996]{flower}
  Flower, P.J. 1996, ApJ 469, 355


\bibitem[Franciosini et al., 2003]{frp03} 
  Franciosini, E., Randich, S., \& Pallavicini, R. 2003, A\&A 405, 551

\bibitem[Giampapa et al., 2000]{grhb00}
  Giampapa, M. S., Radick, R. R., Hall, J. C., \& Baliunas, S. L. 2000, American Astronomical Society, 
  SPD meeting 32, 02/12/2000

\bibitem[Gray, 1992]{gray}
  Gray, D.F. 1992, The observation and analysis of stellar photospheres 
  (Cambridge Astrophphysics Series)

\bibitem[Guenther, 1989]{g89}
  Guenther, D.B. 1989, ApJ 339, 1156

\bibitem[Hempelmann et al., 1996]{hss96}
  Hempelmann, A., Scmhmitt, J.H.M.M., \& St\d{e}pien, K. 1996, A\&A 305, 284

\bibitem[Henry et al., 1996]{hsdb96}
  Henry, T.J., Soderblom, D.R., Donahue, R.A., \& Baliunas, S.L. 1996, AJ 11, 439

\bibitem[Jones \& Cudworth, 1983]{jc83}
  Jones, B.F.,  \& Cudworth, K. 1983, AJ 88, 215

\bibitem[Jones \& Stauffer, 1991]{js91}
  Jones, B.F.,  \& Stauffer, J.R. 1991, AJ 102, 1080

\bibitem[Klein Wassink, 1927]{kw}
  Klein Wassink, W.J. 1927, Publ. Kapteyn Astron. Lab. Groningen 41, 1

\bibitem[La Bonte, 1986]{labonte}
  La Bonte, B.J. 1986,  ApJSS 62, 229

\bibitem[Latham et al., 1992]{lmmd92}
  Latham, D. W., Mathieu, R. D., Milone, A. A. E., \& Davis, R. J. 1992, IAUS 151, 471

\bibitem[Linsky \& Ayres, 1978]{la78}
  Linsky, J. L., \& Ayres, T. R. 1978, ApJ 220, 619

\bibitem[Linsky \& Avrett, 1970]{la70}
  Linsky, J.L., \& Avrett, E.H. 1970, PASP 82, 169

\bibitem[Linsky et al., 1979]{lmrw79}
  Linsky, J.L., McClintock, W., Robertson, R.M., \& Worden, S.P. 1979, ApJS 41, 47

\bibitem[Meibom et al., 2002]{man02} 
  Meibom, S., Andersen, J., \& Nordstr\"{o}m, B. 2002, A\&A 386, 187

\bibitem[Melo et al., 2001]{mpd01}
  Melo, C., Pasquini, L., \& De Medeiros, J.R. 2001, A\&A 375, 851

\bibitem[Mermilliod, 1981]{mermilliod}
  Mermilliod, J.C. 1981, A\&A 97, 235

\bibitem[Montgomery et al., 1993]{mmj93}
  Montgomery, K.A., Marschall, L.A., \& Janes, K.A. 1993, AJ 106, 181

\bibitem[Nordstr\"{o}m et al., 1996]{naa96}
  Nordstr\"{o}m, B., Andersen, J., \& Andersen, M.I. 1996, A\&AS 118, 407

\bibitem[Nordstr\"{o}m et al., 1997] {naa97}
  Nordstr\"{o}m, B., Andersen, J., \& Andersen, M.I. 1997, A\&A 322, 460

\bibitem[Noyes et al., 1984]{nhbdv84} 
  Noyes, R.W., Hartmann, S., Baliunas, S.L., Duncan, D.K., \& Vaughan, A.H. 1984, ApJ 279, 763

\bibitem[Parker, 1970]{parker70}
  Parker, E.N. 1970, ARA\&A 8, 1

\bibitem[Pasquini, 1985]{tesiluca} 
  Pasquini, L. 1985, PHD thesis.

\bibitem[Pasquini et al., 1988]{ppp88}
  Pasquini, L., Pallavicini, R., \& Pakull, M. 1988, A\&A 191, 253

\bibitem[Pasquini et al., 1990]{pbp90}
  Pasquini, L., Brocato, E., \& Pallavicini, R. 1990, A\&A 234, 277

\bibitem[Pasquini, 1992]{luca92} 
  Pasquini, L. 1992, A\&A 266, 347

\bibitem[Pasquini et al., 2003]{przhcn}
  Pasquini, L., Randich, S., Zoccali, M., Hill, V., Charbonnel., C., Nordstr\"{o}m, B. 2003, Detailed Chemical composistion
  of the open Cluster IC 4651: Metals, $\alpha$ elements and Li, in preparation.

\bibitem[Paulson et al., 2002]{psch02}
  Paulson, D. B., Saar, S. H., Cochran, W. D., \& Hatzes, A. 2002, ApJ 124, 572

\bibitem[Paulson et al., 2003]{psc03}
  Paulson, D. B., Snenden, C., \&  Cochran, W. D. 2003, AJ 125, 3185

\bibitem[Perryman et al., 1998]{perryman}
  Perryman, M.A.C., et al., 1998, A\&A 331, 81
%

\bibitem[Radick et al., 1987]{rtldb87}
  Radick, R. R., Thompson, D.T., Lockwood, G.W., Duncan, D.K., \& Baggett, W.E. 1987, ApJ 321, 459

\bibitem[Radick et al., 1998]{rlsb98}
  Radick, R. R., Lockwood, G.W., Stiff, B.A., \&  Baliunas, S.L. 1998, ApJSS 118, 239 

\bibitem[Randich \& Schmitt, 1995]{rs95} 
  Randich, S., \& Scmhmitt, J.H.M.M. 1995, A\&A 298, 115

\bibitem[Rocha Pinto et al., 2000 I]{rmsf00a}
  Rocha Pinto, H.J., Scalo, J., Maciel, W.J., \& Flynn, C.  2000, A\&A 358, 850 (I)

\bibitem[Rocha Pinto et al., 2000 II]{rmsf00b}
  Rocha Pinto, H.J., Scalo, J., Maciel, W.J., \& Flynn, C.  2000, A\&A 358, 869 (II)

\bibitem[Rodrigues et al., 2003]{rlp03}
  Rodrigues, G. R. K., Lyra, W., \& Porto de Mello, G. F. 2003, Boletim da Sociedade Astronomica Brasileira, 23, 131

\bibitem[Sanders, 1977]{sand}
  Sanders, W.L. 1977, A\&AS 27, 89

\bibitem[Skumanich, 1972]{skumanich}
  Skumanich, A. 1972, ApJ 171, 565

\bibitem[Soderblom et al., 1991]{sdj91}
  Soderblom, D.R., Duncan, D.K., \& Johnson, D.R.H. 1991, ApJ 375, 722

\bibitem[Stern et al., 1995]{ssk95}
  Stern, R.A., Schmitt, J.H.M.M., \& Kahabka, P.T. 1995, ApJ 448, 683

\bibitem[Straizis \& Valiuga, 1994]{sv94}
  Straizis, V., \& Valiuga, G. 1994, BaltA 3, 282

\bibitem[Strassmeier et al., 2000]{swgsw2000}
  Strassmeier, K.G., Washuettl, A., Granzer, T., Scheck, M., \& Weber, M. 2000, A\&AS, 142, 275

\bibitem[Taylor \& Joner, 2002]{noext}
  Taylor, B. J.\& Joner, M. D. 2002, AAS 200, 902

\bibitem[van Bueren, 1952]{vanbueren}
  van Bueren, H.G. 1952, Bull. Astron. Inst. Netherlands, 11, 385

\bibitem[van den Bergh \& McClure, 1980]{vBm80}
  van den Bergh, S., McClure, R.D. 1980, A\&A 88, 360

\bibitem[Vaughan \& Preston, 1980]{vp80}
  Vaughan, A.H., \& Preston, G.W. 1980, PASP 92, 385

\bibitem[White \& Livingston, 1981]{wl81} 
  White, O.R., \& Livingston, W. C. 1981, ApJ 249, 798

\bibitem[Wilson, 1963]{w63} 
  Wilson, O.C. 1963, ApJ 138, 832

\end{thebibliography}

\end{document}